\documentclass[10pt]{article}
\usepackage[utf8]{inputenc}
\usepackage{graphicx}
\usepackage{subcaption}
\usepackage{amsthm}
\usepackage{mathrsfs}
\usepackage{slashed}
\begin{document}

\title{\bf Unraveling the Hubble tension with warm inflation}
\date{}
\maketitle
\begin{center}
\large{Anupama B\footnote[1]{21phph19@uohyd.ac.in} and P K Suresh\footnote[2]{sureshpk@uohyd.ac.in (corresponding author)}}\\
\small{School of Physics, University of Hyderabad, Hyderabad, 500046, India.} 
\end{center}
\begin{abstract}
The validity of warm inflation is investigated in the light of recent CMB missions in both strong and weak dissipative regimes. The tensor to scalar ratio of various inflationary models is found to be consistent with the recent CMB results for different models of warm inflation. The role of dissipation on the popular models of warm inflation in the context of supersymmetry and string theory is investigated. Further, the effect of dissipation coefficient of warm inflation on the Hubble parameter and its role in accounting the Hubble tension is examined. Warm inflation embodies superstring theory and can provide a platform to test quantum gravity in multi field scenario.\\
\vspace{1pc}
\end{abstract}
\section{Introduction}
Various recent cosmological observations support large scale structure formation of the universe from primordial fluctuations seeded during inflation \cite{10, planck, 12, par, find}. The inflationary paradigm \cite{guth}, initially proposed to overcome some of the shortcomings of standard cosmology, is still widely accepted among the scientific community to decode the physics of the early universe. It is customary to consider a homogenous scalar field called inflaton as a standard candidate to describe the sudden exponential expansion of the universe. Depending upon the dynamics of the inflaton field,  inflation can be modelled in two ways. The conventional framework where the expansion dilutes the available energy density settling the universe in a cooled state is reffered to as cold inflation (CI) \cite{enc}. Whereas the setup where inflaton interacts with other degrees of freedom during inflation and the associated energy dissipation regulates the temperature of the universe by compensating for the effect of expansion is known as warm inflation (WI) \cite{7}. The warm inflationary scenario can overcome some of the main drawbacks of CI such as the demand for a special reheating phase and the ambiguity on the quantum to classical transition of the fluctuations in the early universe. \\ 
\\
The dynamics of warm inflation depend mainly on dissipative mechanisms. Therefore a suitable model for warm inflation must accomplish a proper channel for sustaining the temperature of the thermal bath. Significant energy dissipation from the inflaton field can be achieved by coupling the inflaton to various thermally excited fields. The large number of couplings can lead to efficient overdamping in the inflaton dynamics, ensuring approximately 60 e-folds. Any decline in the population of decay products must be forbidden during the expansion of the universe in an ideal model of warm inflation. According to warm inflation, the slow roll of inflaton is accompanied by thermal friction from the decay products of dissipation. Therefore, additional couplings can result in more dissipation and overdamping of the inflaton field. Hence, it is interesting to identify those interaction schemes between particle system and heat bath (like Kramer's problem, Caldeira-Leggett model \cite{Cald1, Cald2}) relevant to cosmology. \\
\\
Recently, it has been shown that WI predicts a lower tensor to scalar ratio and hence favours some of the inflationary models that are ruled out in CI \cite{quartic}. WI can naturally accommodate the swampland conjectures \cite{swamp1,swamp2} and can be used to validate the existence of primordial blackholes \cite{bh1, bh2}. These attractive features of WI can be attributed to the various interaction schemes of the inflaton with other degrees of freedom. Supersymmetric interactions in warm inflation suggest the production of dark matter from the decay of inflaton \cite{susydm}. Studies are still in progress towards establishing WI. These aspects make WI a flourishing area with many possibilities to examine inflation. One such possibility is incorporating high energy theories like quantum gravity in warm inflation. Regardless of the efforts, till date, there is no consistent and comprehensive formulation for quantum gravity. Even though it is difficult to observe direct evidences in support of quantum gravity, its effects can be tested indirectly using suitable techniques. Effective field theory (EFT) may be the best available approach to realise this scenario \cite{eft} where the low energy manifestation of higher energy theories (like string theory) can be achieved within a particular cut off. EFTs that are not ultraviolet complete fall in the swampland region \cite{vafa,ooguri}. It is easy to implement quantum gravity in warm inflation by modelling the thermal interactions using string theory, making dissipation intrinsic to warm inflation. These interactions can prevent super-Planckian excursions of the field and make WI effective field theory consistent \cite{eft} thereby alleviating the “$\eta-$problem” prevailing in the string based cold inflationary models \cite{eta}. \\
\\
 Previous studies on warm inflation concentrate only on constructing models that can produce sufficient dissipation to sustain the temperature of the heat bath. Since the interactions contributing to the thermal bath co-occur during warm inflation, it can have an impact on the rate of expansion of the universe as well as on the Hubble parameter. Therefore, the present work primarily aims to explore the role of dissipation on the inflationary dynamics and its importance in analysing the Hubble tension. The cosmological inflationary parameters for different superstring theory motivated warm inflationary models are derived and its compatibility with current and upcoming CMB estimates is examined. A bound on the required strength of dissipation ($Q$) over the effect of expansion of the universe from CMB for renowned models in WI is obtained for different choices of inflaton field and various background fields. The feasibility of weak and strong dissipative regimes with the future projected estimate of tensor to scalar ratio is analysed from various CMB data.  Further, the Hubble tension is addressed in the light of multi field warm inflationary scenario.
\\ 
\section{Cold and Warm inflation}
Consider the inflaton field ($\phi$) and its potential $V(\phi)$ in a flat Friedman-Lemaitre-Robertson-Walker (FLRW) metric.  The Friedmann equation relating the Hubble parameter ($H$) to the total energy density ($\rho$) of the universe is given by,
\begin{eqnarray}
H^2 = \frac{\rho}{3m_{pl}^2}. \label{fried}
\end{eqnarray}
In the cold inflationary scenario, radiation energy density is negligible compared to the vacuum energy density of inflaton. At the end of inflation, the inflaton decays into standard model particles through parametric resonance followed by a reheating and radiation era.
The equation of motion of the inflaton in this case is,
\begin{eqnarray}
\ddot \phi + 3H \dot \phi + V'(\phi) &=& 0. 
\end{eqnarray}
The prime $(')$ and dot ($.$)  indicates derivative with respect to the field ($\phi$) and time ($t$) respectively. 
Assuming that the inflaton rolls slowly ($\frac{\dot\phi^2}{2}<<V(\phi)$) towards the minimum of its potential, the slow roll parameters are defined as
\begin{eqnarray}
\epsilon_H&=& \frac{-\dot H}{H^2} \simeq \frac{m_{pl}^2}{2}\bigg(\frac{V'(\phi)}{V(\phi)}\bigg)^2 \simeq \epsilon_v, \\ 
\eta_H &=& \frac{- \ddot \phi}{H \dot \phi} \simeq m_{pl}^2 \frac{V''(\phi)}{V(\phi)} \simeq \eta_v.
\end{eqnarray}
The power spectrum corresponding to the evolution of primordial scalar and tensor perturbation in the cold inflationary scenario is\\
\begin{eqnarray}
P_s^{CI}(k)&=& A_s(k_0) \  \bigg(\frac{k}{k_0}\bigg)^{n_s^{CI} -1},\\
P_t^{CI}(k)&=& A_t(k_0) \  \bigg(\frac{k}{k_0}\bigg)^{n_t^{CI}},
\end{eqnarray}
where $A_s({k_0}) =\frac{1}{12\pi^2m_{pl}^2}\frac{V(\phi)^3}{V(\phi)'^2} \bigg|_{k=k_0}$ and $A_t({k_0}) =  \frac{2}{3\pi^2m_{pl}^4}V(\phi) \bigg|_{k=k_0}.$
The ratio of the tensor power spectrum to that of the scalar power spectrum evaluated at horizon crossing is called the tensor to scalar ratio,
\begin{eqnarray}
r^{CI}  = \frac{P_t^{CI}(k_0)}{P_s^{CI}(k_0)} = 16\epsilon_v. \label{eq5}
\end{eqnarray}
The scale dependency of the power spectra can be respectively expressed in terms of the corresponding spectral indices,
\begin{eqnarray}
n_{s}^{CI}  -1  = \frac{\mathrm{d} \ \mathrm{ln}P_s^{CI}}{\mathrm{d} \  \mathrm{ln}k}
\end{eqnarray}
and
\begin{eqnarray}
n_{t}^{CI}  = \frac{\mathrm{d} \ \mathrm{ln}P_t^{CI}}{\mathrm{d} \ \mathrm{ln}k}.
\end{eqnarray}
The scalar spectral index can be expressed in terms of the slow roll parameters
\begin{eqnarray}\label{ns}
n_s^{CI}  = 1 + 2\eta_v  - 6\epsilon_v.
\end{eqnarray}
Based on various particle physics and string theory interactions, more than hundreds of models have been hitherto proposed to describe inflation (for a review of recent models see \cite{enc}). A systematic approach is employed to validate these models by comparing its predictions with various CMB observations. The advancements in precision cosmology in the last decade have tremendously improved the quality of observational data. The joint data analysis from Planck 2018 and BK18 suggests a bound on the tensor to scalar ratio ($r < 0.036$) \cite{10} which rules out many popular models in cold inflationary scenario.\\
\\
The inflationary potential can take simplest forms like 
\begin{eqnarray}
V(\phi) = \cases{
            \frac{1}{2} m^2 \phi^2, & Quadratic chaotic inflation \cr
            \frac{\lambda}{4}\phi^4,&  Quartic chaotic inflation. \cr
            }\label{Val}
\end{eqnarray}
The cosmological inflationary parameters of such potentials (\ref{Val}) in the conventional cold inflationary setting is given in Table(\ref{coldcosm}).
\begin{table}
\caption{\label{coldcosm}Cosmological inflationary parameters for quadratic and quartic cold inflationary model in terms of the e-folding number ($N$).}
\begin{tabular*}{0.8\textwidth}{@{}l*{15}{@{\extracolsep{0pt plus12pt}}l}}
\hline
Inflationary parameter & $V(\phi)=\frac{1}{2}m^2\phi^2$ & $V(\phi)=\frac{\lambda}{4}\phi^4$ \\
\hline
$\epsilon_v$  & $\frac{1}{2N}$ & $\frac{1}{N}$ \\
$\eta_v$ & $\frac{1}{2N}$ & $\frac{3}{2N}$  \\
$n_s^{CI}$ & $1-\frac{2}{N}$ & $1-\frac{3}{N}$  \\
$r^{CI}$  & $\frac{8}{N}$ & $\frac{16}{N}$ \\
\hline
\end{tabular*}
\end{table}
For $N=60$, 
the scalar spectral index and tensor to scalar ratio of these models do not fall in the observational limit of recent CMB data \cite{10} and hence these models are disfavoured in CI. Moreover, CI could not satisfactorily explain some of the issues like the transition of fluctuations from quantum to classical. In CI, it is necessary to have an additional preheating phase with a distinct mechanism to reheat the universe which is still not understood completely. This problem can be overcome by an alternative scenario known as warm inflation, 
where a non vanishing radiation bath can act as thermal seed for density perturbation vital for the large scale structure formation. Since the thermal fluctuations are classical in origin, warm inflation need not explain the quantum to classical transition of primordial perturbations that is otherwise present in cold inflation.\\ 
\\ 
The fundamental requirement of warm inflation is the coupling of inflaton to other fields while remaining in contact with the thermal bath. The generic form of the Lagrangian density in a warm inflationary scenario that satisfies these conditions can be written as \cite{silva},
\begin{eqnarray}
\mathcal{L}[\phi, X, Y ] =& \mathcal{L}[\phi] + \mathcal{L}[X] + \mathcal{L}[Y ] +  \mathcal{L}_{int}[\phi, X] + \mathcal{L}_{int}[X, Y ] 
\end{eqnarray}\label{warmlag}
where $\phi$ is the inflaton field, $X$ is any field(s) that can be coupled directly to the inflaton field and $Y$ denotes field(s) that may not be directly coupled to the inflaton, but via $X$. $\mathcal{L}_{int}[\phi, X] $ and $\mathcal{L}_{int}[X, Y ] $ give the interaction among these fields. The energy dissipated into the bath as the inflaton decays into lighter particles and attains the lowest energy state is given by the dissipation coefficient ($\gamma$). The coefficient of dissipation arising from such models has a generic form \cite{kamal},
\begin{eqnarray}
\gamma(\phi,T) = A_{\gamma}T^l \phi^p M^{1-p-l}.
\end{eqnarray}
where $l,p$ are dimensionless integers, $A_\gamma$ is a dimensionless constant that carries the details of the microscopic model and $M$ is the cut off mass scale of the particular model under consideration. At the end of inflation, this leads to a smooth transition of the universe into a radiation dominated era without the requirement of a separate reheating phase. The decay products generate an effective viscosity ($\gamma \dot \phi$) in the thermal bath and this can be incorporated along with the Hubble friction ($H\dot\phi$) during expansion in the equation of motion \cite{8},
\begin{eqnarray}
\ddot \phi + 3H \dot \phi +  \gamma \dot \phi + V'(\phi) &=&0, \\ 
\ddot \phi + 3H (1+Q) \dot \phi + V'(\phi) &=& 0, \label{motion}\\
\dot \rho_r + 4H \rho_r &=&  \gamma \dot \phi^2 .\label{cont}
\end{eqnarray}
Here, $Q = \frac{\gamma}{3H}$ represents the thermal equilibrium between the rate of production of radiation due to the interaction and its rate of dilution due to expansion.  The dissipation can be either strong ($Q>>1$) or weak ($Q<<1$) depending on the nature of interaction. Note that when $Q=0$, the cold inflationary case can be recovered.  \\ \\
The associated radiation energy density  ($\rho_r$) of the thermal bath with a temperature ($T$) is,
\begin{eqnarray}
\rho_r = \frac{\pi^2}{30}g_*T^4 \label{rad}
\end{eqnarray} where $g_*$ is the effective degrees of freedom available in the bath. During warm inflation, it is crucial to maintain the condition $\rho_r^{\frac{1}{4}} > H$ \cite{7,kaushik} 
and it is also essential that the thermal bath must attain a steady state ($\dot \rho_r =0$). Therefore from (\ref{cont}),
\begin{eqnarray}
\rho_r = \frac{\gamma \dot \phi^2}{4H} = \frac{3}{4}Q\dot \phi^2 \label{ror}.
\end{eqnarray}
Applying the slow roll condition, $\frac{\dot \phi^2}{2} << V(\phi)$, on the warm inflation, the equation of motion
(\ref{motion}) gets modified,
\begin{eqnarray}
3H(1+Q)\dot \phi = -V'(\phi). \label{slowmotion}
\end{eqnarray}
and the corresponding first and second slow roll parameters are respectively,
\begin{eqnarray}
\epsilon_H &= & \frac{\epsilon_v}{1+Q}, \label{epwarm} \\ 
\eta_H &=&  \frac{1}{1+Q}(\eta_v - \beta + \frac{\beta -\eta_v}{1+Q}), 
\end{eqnarray}
where, 
\begin{eqnarray}
\beta = \frac{\gamma' V'(\phi)}{\gamma V(\phi)}. \label{bet}
\end{eqnarray}
The equation for evolution of the thermal fluctuations ($\delta \phi$) is given by,
\begin{eqnarray}
\delta \ddot  \phi(k,t) + (3H+ \gamma) \delta \dot  \phi(k,t) +  (\frac{k}{a^2}+m^2)\delta\phi(k,t) = \xi(k,t)
\end{eqnarray}
where, $\xi$ is the thermal noise fluctuation and is related to the dissipation through the fluctuation-dissipation relation \cite{fluctdiss}. Hence its average over the statistical ensemble is
\begin{eqnarray}
<\xi(x,t)\xi(x',t)> \ = 2\gamma T a^{-3} \delta(x-x')\delta(t-t').
\end{eqnarray}
The variance of thermal fluctuations can be expressed as \cite{fluct},
\begin{eqnarray}
<\delta \phi^2> \ \simeq \frac{k_F T}{2\pi^2}.
\end{eqnarray}
Where $k_F$ is the wavenumber of thermal fluctuations at freeze out. In the limit $Q>>1$, $k_F \sim \sqrt{H\gamma}$ and for $Q<<1$, $k_F \sim H$.\\
\\The scalar power spectrum associated with the thermal fluctuations of the warm inflation, where $\gamma$ is independent of temperature ($l=0$) is given by \cite{silva},
\begin{eqnarray}
P_{s_{l=0}}^{WI} &=& \bigg(\frac{H^2}{2\pi \dot\phi} \bigg)^2 \bigg[\coth(\frac{H}{2T}) + \frac{T}{H} \frac{2\sqrt{3}\pi Q}{\sqrt{3+4\pi Q}}\bigg].\label{ps}
\end{eqnarray} 
If $\gamma$ is coupled to temperature ($l \neq 0$), then the scalar power spectrum gets modified to
\begin{eqnarray}
P_{s_{l \neq 0}}^{WI} &=& P_{s_{l=0}}^{WI} \times G(Q),
\end{eqnarray}
where,
\begin{eqnarray}
G_{l=3}(Q)&=&  1+4.981Q^{1.946}+0.127Q^{4.336},\\
G_{l=1}(Q)  &=&  1+0.335Q^{1.364}+0.0185Q^{2.315},\\
G_{l=-1}(Q) &=& \frac{1+0.4Q^{0.77}}{(1+Q^{1.09})^2}. \label{Gq}
\end{eqnarray}

$\frac{T}{H}$ in the scalar power spectrum (\ref{ps}) can be expressed in terms of $Q$ from (\ref{fried}), (\ref{rad}) , (\ref{ror}) and (\ref{slowmotion})  as,
\begin{equation}\label{tH}
\frac{T}{H} =   c \bigg[ \frac{Q}{(1+Q)^2} \bigg]^{\frac{1}{4}},
\end{equation}
where 
\begin{eqnarray}
c &=& \bigg( \frac{135}{64} \frac{\epsilon_v}{\pi^4 V g_*} \bigg)^{\frac{1}{4}} \label{tch}.
\end{eqnarray}
The thermal interactions do not contribute to the tensor perturbations therefore the tensor power spectrum remains the same as that of CI, 
\begin{eqnarray}
P_t^{WI} &=& P_t^{CI} = \bigg( \frac{H^2}{\sqrt{2} \pi m_{pl}^2} \bigg)^2.
\end{eqnarray}
Therefore the corresponding scalar spectral index and the tensor to scalar ratio become \cite{oyv},
\begin{eqnarray}
1-n_s ^{WI} &=& \frac{1}{1+Q} \bigg\{4\epsilon_v - 2 \bigg( \eta_v - \beta + \frac{ \beta - \epsilon_v}{1+Q} \bigg) +  \frac{\alpha}{1+\alpha} \bigg[\frac{2\eta_v+\beta-7\epsilon_v}{4} +  \frac{6+(3+4\pi)Q}{(1+Q)(3+4\pi Q)} \ (\beta-\epsilon_v)\bigg] \bigg\} \\&& \nonumber- \frac{\mathrm{d}G(Q)}{\mathrm{d} \ \mathrm{ln}k} , \\
r^{WI} &=& \frac{16 \epsilon_v}{[\coth( \frac{H}{2T})+ \alpha]G(Q)}, \label{rwarm}
\end{eqnarray}
where, 
\begin{eqnarray}
\alpha =  \frac{T}{H}\frac{2\sqrt{3}\pi Q}{\sqrt{3+4\pi Q}},
\end{eqnarray} and 
\begin{eqnarray}
\coth( \frac{H}{2T})=1+ 2n_k.
\end{eqnarray} 
Here, $n_k$ is the distribution of inflaton particles in the thermal bath. For CI, $n_k =0$ and for WI it takes Bose-Einstein distribution \cite{RR},
  \begin{eqnarray}
n_k = \frac{1}{\exp(\frac{k}{aT})-1}.
 \end{eqnarray} 
In the strong dissipation regime ($Q>>1$ and $\alpha>>1$), we get the scalar spectral index for the WI as,
\begin{eqnarray}
n_s ^{WI} &=&1 - \frac{3}{2Q} \bigg[ \frac{3}{2} (\epsilon_v+\beta)-\eta_v \bigg] +  2Q^{2} \bigg(1+ \frac{\epsilon_v}{1+Q} \bigg)  \frac{(2+l) \epsilon_v - l \eta_v - 2p \frac{\epsilon_v}{\phi}}{4-l+(4+l)Q}  \frac{1}{G(Q)} \frac{\mathrm{d}G(Q)}{\mathrm{d \ ln}k}
\end{eqnarray}
and  the tensor to scalar ratio,\begin{eqnarray}
r^{WI} &=& \frac{16 \epsilon_v}{\sqrt{3\pi}cQ^{\frac{1}{4}}G(Q)}.
\end{eqnarray}\\
Similarly, in the weak dissipation regime ($Q<<1$, $\alpha<<1$, $G(Q)\simeq 1$ and $\frac{\mathrm{d}G(Q)}{\mathrm{d \ ln}k} = 0$), we get the corresponding scalar spectral index
\begin{eqnarray}
n_s ^{WI} &=& 1 - [2(3\epsilon_v-\eta_v)-\frac{\alpha}{4} (15\epsilon_v -2\eta_v-9\beta)], 
\end{eqnarray}
and tensor to scalar ratio
 \begin{eqnarray}
r^{WI} &=& \frac{8\epsilon_v}{cQ^{\frac{1}{4}}}.
\end{eqnarray}\\

\section{Warm inflationary models} 
Warm inflationary model building is based on different types of interaction between inflaton and other environmental fields \cite{sdb}. The available literatures focus only on various methods to derive the dissipation coefficient and to enhance the dissipation. These qualitative studies have facilitated the construction of better models of warm inflation over the years. So far, four major models of warm inflation have been proposed inspired from supersymmetry and string theory \cite{story, dev}. In this section, we discuss the microphysical framework of these models and the challenges involved in implementing them.  We attempt to study warm inflation quantitatively from the existing and forthcoming CMB estimates by constraining the factor $Q$ that gives the dominance of dissipation over the effect of expansion. We also investigate the relevance of weak and strong dissipative regimes for various warm inflationary models with the latest and prospected CMB estimates and its implications in validating warm inflation and inflationary model.

\subsection{Distributed mass model (DMM)}
It is a string theory inspired model of warm inflation where the inflaton couples to a spectrum of string modes \cite{ABTK, gill}.   Phenomenological realisation of the higher levels of the strings can be achieved using effective theory where the wide range of vacua in string theory \cite{doug} can automatically give rise to large number of fields $(>10^4)$. In the low energy realisation, inflaton can interact with this different tower of states of the string. Inflaton can exhibit a shifted interaction with these various states as it scans the tower that is crucial for dissipation. This model has an inherent landscape of energy levels arising from the properties of string vacuum \cite{kep}.\\
\\
We consider a toy model exhibiting $\mathcal{N}=1$ global supersymmetry with the inflaton  ($\phi$) coupled to massive modes of the string. The corresponding Lagrangian for the interaction of inflaton with $N_M \times N_\chi$ scalar fields ($\chi_{jk}$) and $N_M \times N_\psi$ fermionic fields ($\psi_{jk}$) can be expressed as \cite{disquo},
\begin{eqnarray}
\mathcal{L}[\phi,\chi_{jk}, \bar \psi_{jk},\psi_{jk}] &= \frac{1}{2}(\partial_\mu \phi)^2 - V(\phi) + \sum_{j=1}^{N_M} \sum_{k=1}^{N_\chi} { \frac{1}{2} (\partial_\mu \chi_{jk})^2 - \frac{f_{jk}}{4!} \chi_{jk}^{4} -  \frac{g_{jk}^{2}}{2} ( \phi - M_{j})^2 \chi_{jk}^{2} } \\& \nonumber +  \sum_{j=1}^{N_M} \sum_{k=1}^{\frac{N_\chi}{4}} { i \bar \psi_{jk} \not{\partial} \psi_{jk} - h_{jk}(\phi - M_j) \bar \psi_{jk} \psi_{jk}}. \label{lagdmm}
\end{eqnarray}
Here, $\lambda$ and $f_{jk}$=$g_{jk}$=$h_{jk}$=$g$ are the various coupling constants. As the inflaton rolls down its potential, the interaction of $\phi$ with $\chi_{jk}$ and $\psi_{jk}$ fields establishes a mass scale distribution $g^2(\phi-M_j)^2$. The effective particle degrees produced in the interaction is given by,
\begin{eqnarray}
g_{*} = 2N_{ \chi} + 7N_{ \psi} +1.
\end{eqnarray}
In DMM, the dissipation has a contribution from both bosonic and fermionic fields, i.e, $\gamma = \gamma^{(B)}+\gamma^{(F)}$. In terms of the equilibrium number densities ($n^{eq}$), the field dependent dissipation is given by \cite{disquo},
\begin{eqnarray}\label{DMMdis}
\gamma^{(B)}(\phi) &=& \frac{\lambda^2\phi^2}{8T} \int \frac{d^3q}{(2\pi)^3} \frac{n_{\phi}^{eq} (1+n_{\phi}^{eq})}{\omega_{\phi}^2(q)\Gamma_{\phi}(q)}  + \sum_{i=1}^{N_M} \sum_{j=1}^{N_\chi} \frac{g^4(\phi-M_i)^2}{T} \int \frac{d^3q}{(2\pi)^3} \frac{n_{\chi}^{eq} (1+n_{\chi}^{eq})}{\omega_{\chi}^2(q)\Gamma_{\chi}(q)}, \\ \label{dis2}
\gamma^{(F)}(\phi)& =& \sum_{i=1}^{N_M}\sum_{j=1}^{\frac{N_\chi}{4}} \frac{g^2(\phi-M_i)^2}{T} \times  \int \frac{d^3q}{(2\pi)^3} \frac{n_{\psi}^{eq} (1+n_{\psi}^{eq})}{\omega_{\psi}^2(q)\Gamma_{\psi}(q)}.
\end{eqnarray}
For a given temperature $T$, only the fields with masses $ g^2(\phi-M_i)^2 << T^2$ will contribute to the thermal viscosity. Equation (\ref{DMMdis}) and (\ref{dis2}) suggests that the number of mass scales ($M_i$) can have a direct impact on dissipation dynamics and duration of warm inflation. Therefore the DMM must exhibit strong dissipation with $Q>>1$.
Note from (\ref{DMMdis}) and  (\ref{dis2}) that the damping coefficient depends on $ g^2(\phi-M_i)^2$ or $\phi^2$. From (\ref{bet}) and (\ref{Gq}), for $\gamma(\phi) \propto  \frac{\phi^2}{T}$, $\beta = \frac{1}{N}$ and $G(Q)=G_{l=-1}(Q)$. To assess the compatibility of DMM and to explore the possibility of constraining the value of Q, we study the cosmological inflationary parameters in warm inflation.
\begin{table}
\caption{\label{dmmslow}
Cosmological inflationary parameters of quadratic and quartic warm inflation for DMM in terms of the e-folding number ($N$) and $Q$.} 
\begin{tabular*}{0.8\textwidth}{@{}l*{15}{@{\extracolsep{0pt plus12pt}}l}}
\hline
Inflationary   & $V(\phi)=\frac{1}{2}m^2\phi^2$ & $V(\phi)=\frac{\lambda}{4}\phi^4$ \\
parameter & & \\
\hline
$\epsilon_H$  & $\frac{1}{2NQ}$ & $\frac{1}{NQ}$  \\ \\
$\eta_H$ & $-\frac{1}{2NQ}$ & $\frac{1}{2NQ}$  \\ \\
$n_s^{WI}$ & $1- \frac{21}{8NQ}$ & $1- \frac{9}{4NQ} $  \\ \\
& $- \frac{136 Q^2  G'_{l=-1}(Q)}{N(5+3Q)G_{l=-1}(Q)}$& $- \frac{542 Q^2  G'_{l=-1}(Q)}{N(5+3Q)G_{l=-1}(Q)}$\\ \\
$r^{WI}$  & $\frac{8}{ \sqrt{3\pi}NcQ^\frac{1}{4}G_{l=-1}(Q)} $ & $\frac{16}{ \sqrt{3\pi}NcQ^\frac{1}{4}G_{l=-1}(Q)} $ \\
\hline
\end{tabular*}
\end{table}
\begin{figure}[h!]
\centering
\makebox[\linewidth]
{
   \begin{subfigure}{0.5\linewidth}
    \includegraphics[width=\linewidth]{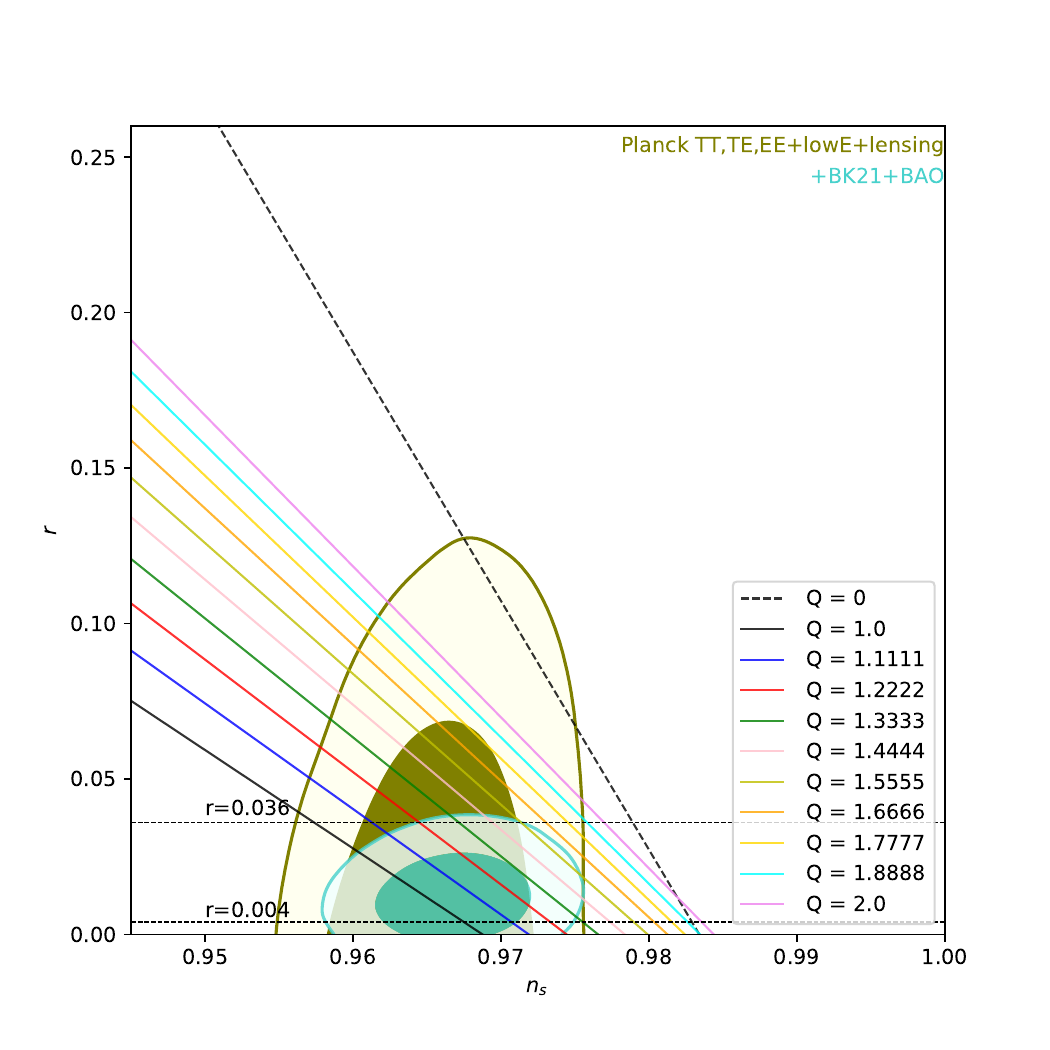}
     \caption{Quadratic}
     \end{subfigure}
     \begin{subfigure}{0.5\linewidth}
    \includegraphics[width=\linewidth]{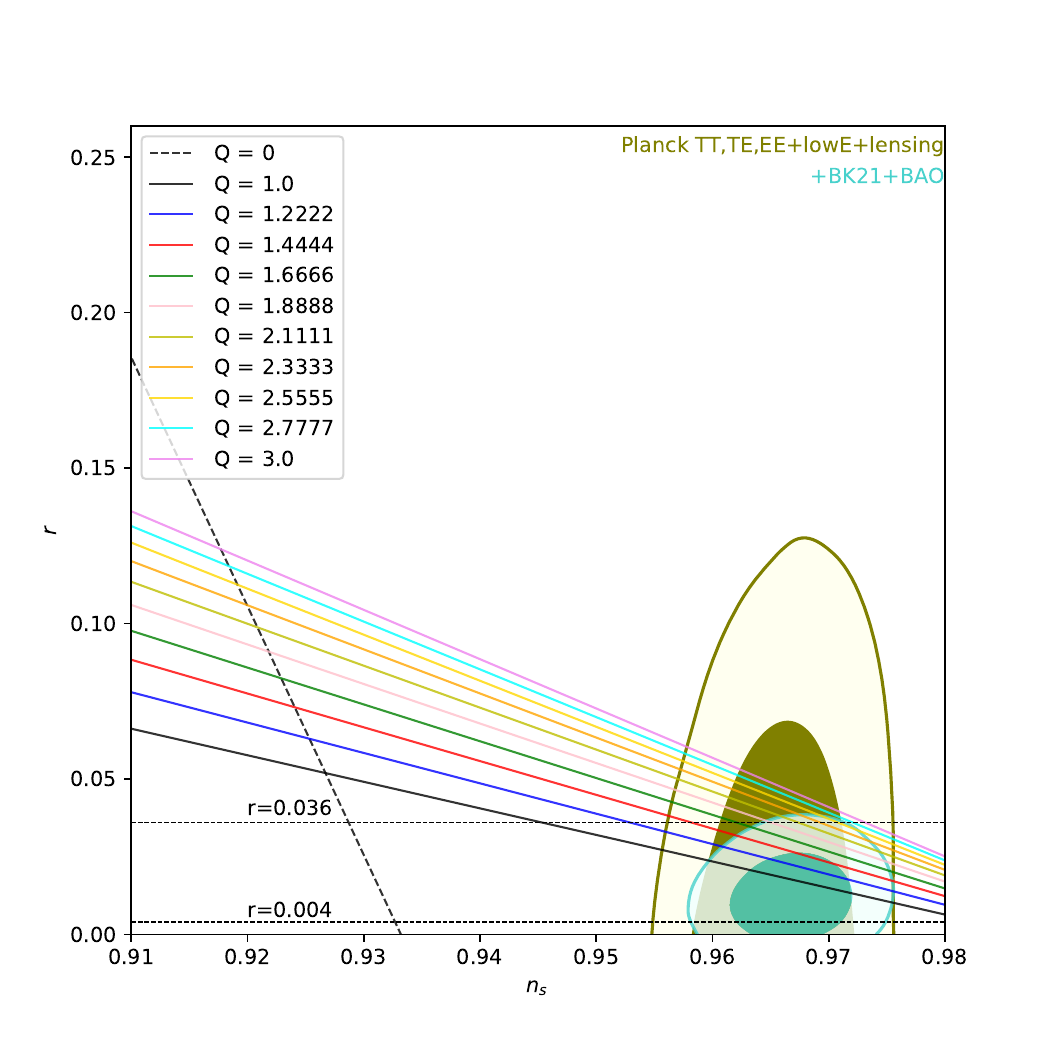}
    \caption{Quartic}
     \end{subfigure}
}
\caption{\label{rnsdmm}Effect of dissipation on the tensor to scalar ratio $(r)$ of quadratic and quartic inflation in DMM for a range of scalar spectral index ($n_s$) with 95$\%$ and 68$\%$ CL of joint Planck TT,TE,EE+lowE+lensing and BK18+BAO.}
\end{figure}
The results of the study are presented in Table(\ref{dmmslow}) and Figure(\ref{rnsdmm}), where $ G'_{l=-1}(Q)=\frac{\mathrm{d}G_{l=-1}(Q)}{\mathrm{d }Q}= \frac{0.3 Q^{-0.23}+0.3Q^{0.86}-2Q^{0.09}}{(1+Q^{1.09})^3}$. The obtained results for $N=60$ show that in DMM there is a drastic decrease in the tensor to scalar ratio of quadratic and quartic inflation. These results are consistent with those reported by joint Planck18 and BK18 observation \cite{10} and therefore this model is viable for warm inflation. We also found an allowed range $1<Q<1.3$ for quadratic inflation and $1<Q<1.4$ for quartic inflation from the 68\% CL of joint Planck18 and BK18. The future projected estimates of $n_s$ and $r$ from LiteBIRD \cite{hazumi} and CMB S4 \cite{s4} permits  $1<Q<1.1$ and $1<Q<1.4$ for quadratic inflation from 68\% CL and 95\% CL of joint Planck18 and BK18 respectively. But the forecast does not favour quartic inflation in DMM. Even though the predictions of DMM align well with the constraints from CMB, the direct coupling of inflaton to the background fields will lead to large thermal corrections affecting the renormalisability of this model. The mass distribution function can avoid thermal corrections to the inflaton mass. For minimizing the radiative corrections,  $N_{\psi}=\frac{N_{\chi}}{4}$ is preferred \cite{dj}. 

\subsection{Two stage mechanism (TSM)}
\begin{figure}[h!]
\centering
\makebox[\linewidth]
{
   \begin{subfigure}{0.5\linewidth}
    \includegraphics[width=\linewidth]{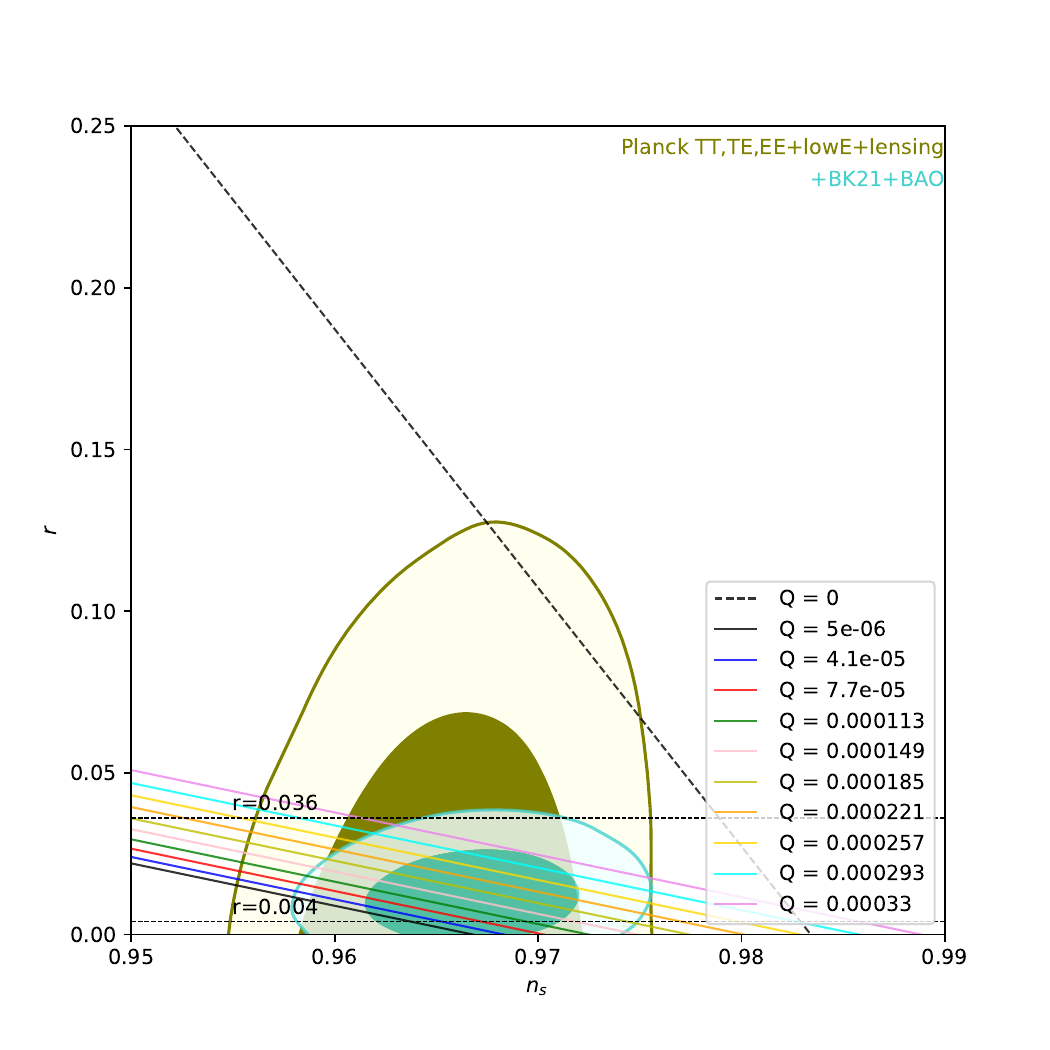}
     \caption{Quadratic}
     \end{subfigure}
     \begin{subfigure}{0.5\linewidth}
    \includegraphics[width=\linewidth]{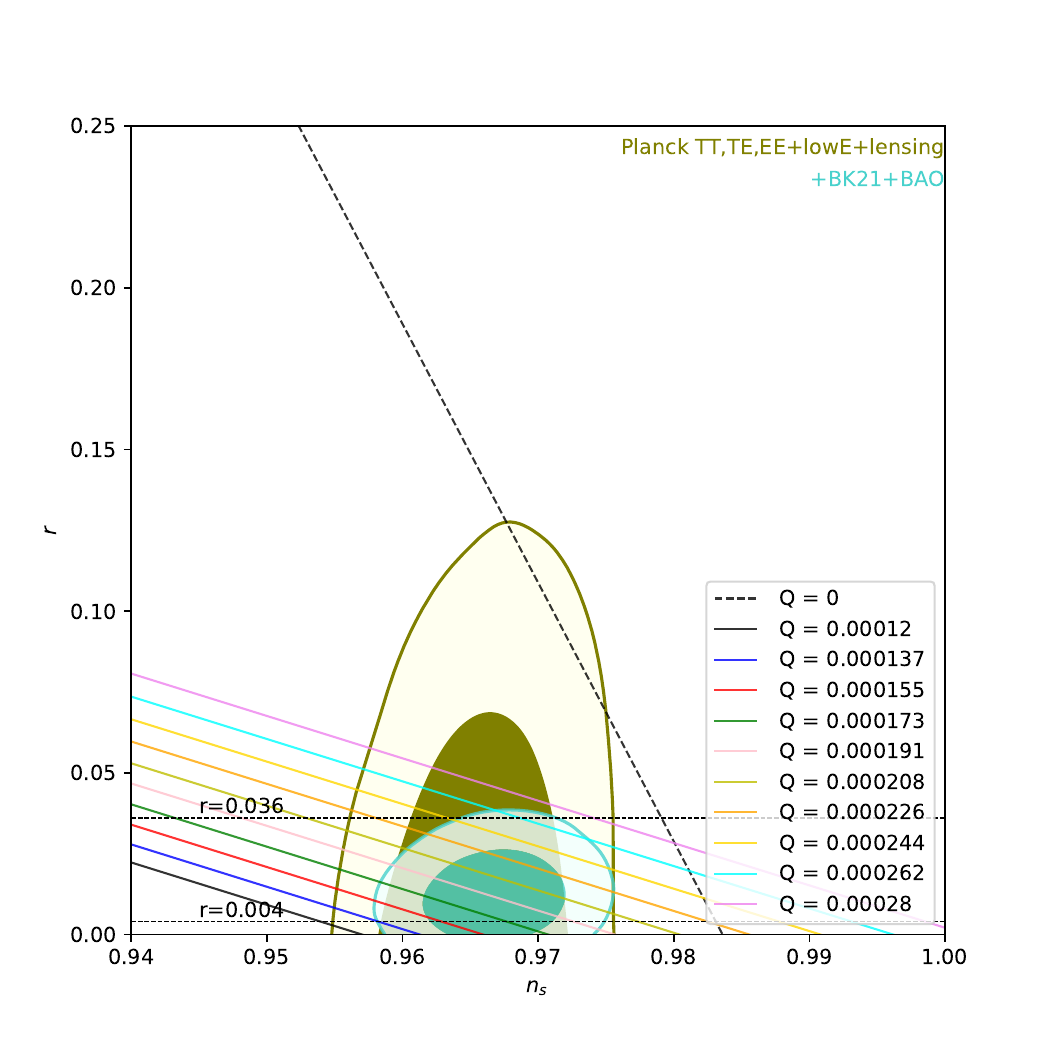}
    \caption{Quartic}
     \end{subfigure}   
}
\caption{\label{rnstwoquad}Effect of dissipation on the tensor to scalar ratio $(r)$ of quadratic and quartic TSM model for a range of scalar spectral index ($n_s$) with 95$\%$ and 68$\%$ CL of joint Planck TT,TE,EE+lowE+lensing and BK18+BAO.}
\end{figure}

In two stage mechanism model of warm inflation, inflaton does not couple directly to the radiation fields, but instead to heavy intermediate fields, which can be either bosons ($\chi$ or $\sigma$) or fermions ($\psi_{\chi}$, $\psi_{\sigma}$). These heavy fields can, in turn, interact with the lighter radiation fields, which remain disconnected from the inflaton sector. These processes can change the masses of heavy fields through which inflaton can dissipate energy into the bath, which then decay into the light radiation fields. This can be executed through a supersymmetric model that contains all the ingredients of a hybrid inflationary model where it is the waterfall field that interacts and decays into lighter degrees of freedom. The superpotential can be written as \cite{mast},
\begin{eqnarray}
W = g \Phi X^2 + h XY^2.
\end{eqnarray}
Here, $\Phi = (\phi,\psi_{\phi})$ represents the inflaton, $X =(\chi,\psi_{\chi})$ is a set of heavy catalyst field and $Y=(\sigma, \psi_{\sigma})$ is a set of lighter radiation field. At tree level, $\phi$ is only coupled to either the scalar boson $\chi$ or fermion $\psi_\chi$. Coupling between scalar field and light fermionic field can be easily forbidden by imposing an $U(1)$ global supersymmetry. SUSY can control potentially large radiative corrections to the inflaton. Such interactions arise naturally in D-brane constructions and are implemented using related gauge theories.\\
\\
The scalar Lagrangian for the aforementioned interactions can be written as \cite{mast,TSM},
\begin{eqnarray}
\mathcal{L}_{int} = &g^2 |\chi|^4 +h^2 |\sigma|^4 +   4g^2 |\chi|^2 |\phi|^2 x +  4gh Re[\phi^{\dagger} \chi^{\dagger} \sigma^2] +  4h^2 |\chi|^2 |\sigma|^2.
\end{eqnarray}
We assume that the couplings $g$ and $h$ are real as complex coupling may generate baryon asymmetry through dissipation. Large values of number multiplets $N_{X,Y}$ of the fields are required to obtain sufficient dissipation in these models. Usually, the Kaluza-Klein tower in extra dimensional scenarios can allow large multiplicities. Supersymmetry breaking can lead to low momentum modes (LM) where the fermionic decays are negligible as well as on shell modes where bosonic and fermionic branching ratios are identical. We focus on the off-shell modes \cite{TSM} with dissipation coefficient
\begin{eqnarray}
\gamma^{LM} \simeq c_{\phi}\frac{T^3}{\phi^2},
\end{eqnarray}
where, $c_{\phi} = 0.02h^2N_Y$. Since $\gamma \propto \frac{T^3}{\phi^2}$, (\ref{bet}) gives $\beta = \frac{-1}{ N}-3(1+Q)$ and $G_{l=3}(Q) \simeq 6.108$ as $Q<<1$. Using this relation, we compute the inflationary parameters for quadratic and quartic model of inflation in the warm inflationary scenario and the results are presented in Table(\ref{tsmslow}) and Figure(\ref{rnstwoquad}).  From the analysis we show that the newly obtained values of scalar spectral index and reduced values of tensor to scalar ratio for $N=60$ are consistent with the joint Planck TT,TE,EE+lowE+lensing and BK18+BAO data with the marginalized 68\% CL. Further, we obtain a bound $5 \times 10^{-6}<Q<0.0003$ and $0.00015<Q<0.0002$ for quadratic and quartic inflation respectively in the TSM model from the 68\% CL of joint analysis of Planck18 and BK18. The upcoming CMB observations support $0<Q<0.000113$ for quadratic inflation and $0.000155<Q<0.00019$ for quartic inflation with 68\% CL  respectively in TSM model of warm inflation from the joint data analysis of Planck18 and BK18. Since we obtain $Q<<1$, this model belongs to the weak dissipative regime. Two stage mechanism fails to achieve strong dissipation, which is one of the main disadvantage of this model. 

\begin{table*}
\centering
\caption{\label{tsmslow}
Cosmological inflationary parameters of quadratic and quartic warm inflation for TSM in terms of the e-folding number ($N$) and $Q$.}
\begin{tabular*}{\textwidth}{@{}l*{15}{@{\extracolsep{0pt plus12pt}}l}}
\hline
Inflationary parameter & $V(\phi)=\frac{1}{2}m^2\phi^2$ & $V(\phi)=\frac{\lambda}{4}\phi^4$ \\
\hline
$\epsilon_H$  & $ \frac{1}{2N(1+Q)}$ & $\frac{1}{N(1+Q)}$ \\
$\eta_H$ & $3Q+\frac{3Q}{2N}$ & $\frac{5Q}{2N}+3Q$  \\
$n_s^{WI}$ & $1- \frac{3}{N}+\frac{\pi cQ^{\frac{5}{2}}}{2}\bigg(\frac{31}{2N}+27 \bigg)$ & $1- \frac{6}{N}+\frac{\pi c Q^\frac{5}{4}}{2}\bigg(\frac{21}{N}+27\bigg)$ \\
$r^{WI}$  & $\frac{4}{ NcQ^\frac{1}{4}G_{l=3}(Q)} $ & $\frac{8}{ NcQ^\frac{1}{4}G_{l=3}(Q)}  $\\
\hline
\end{tabular*}
\end{table*}

\subsection{Warm little inflation (WLI)}
\begin{figure}[h!]
\centering
\makebox[\linewidth]
{ \begin{subfigure}{0.5\linewidth}
    \includegraphics[width=\linewidth]{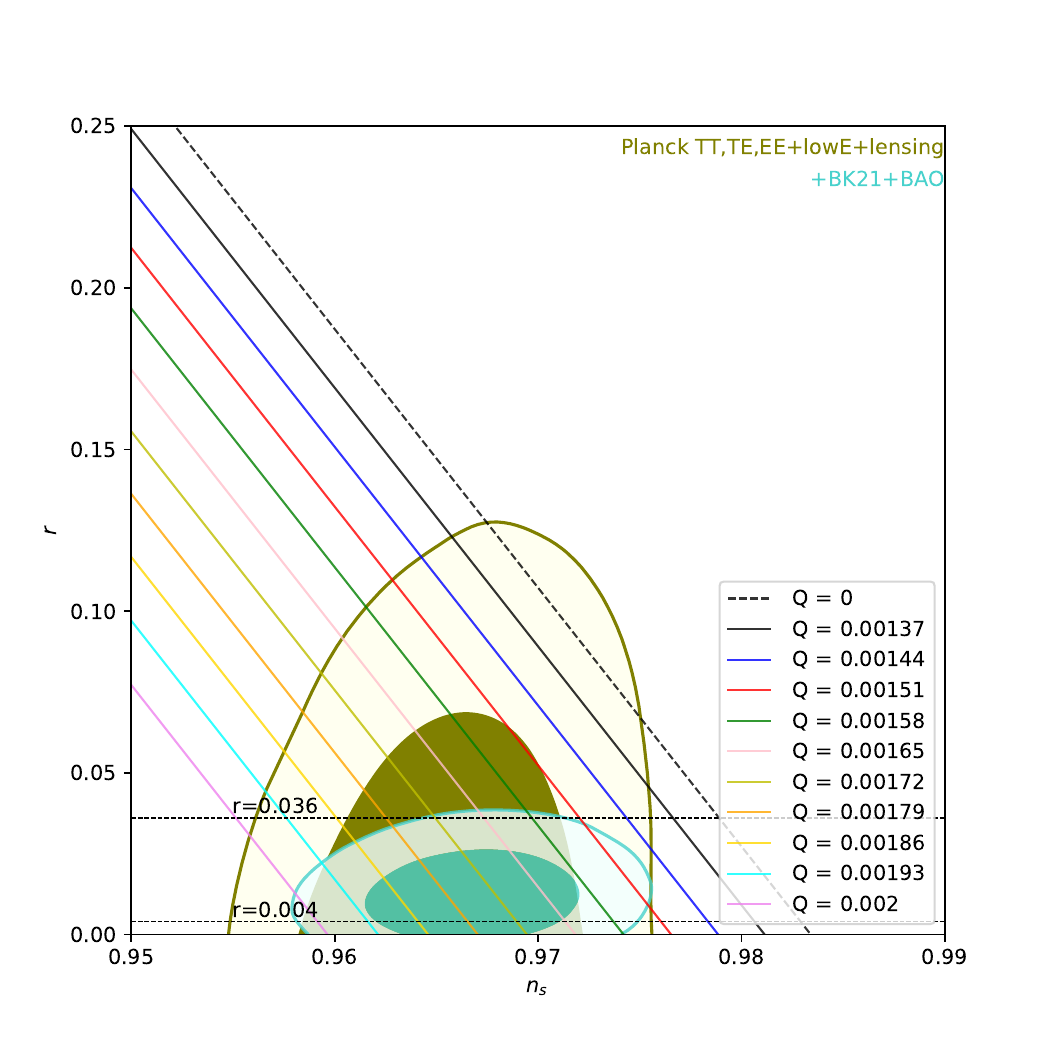}
     \caption{Quadratic}
     \end{subfigure}
     \begin{subfigure}{0.5\linewidth}
    \includegraphics[width=\linewidth]{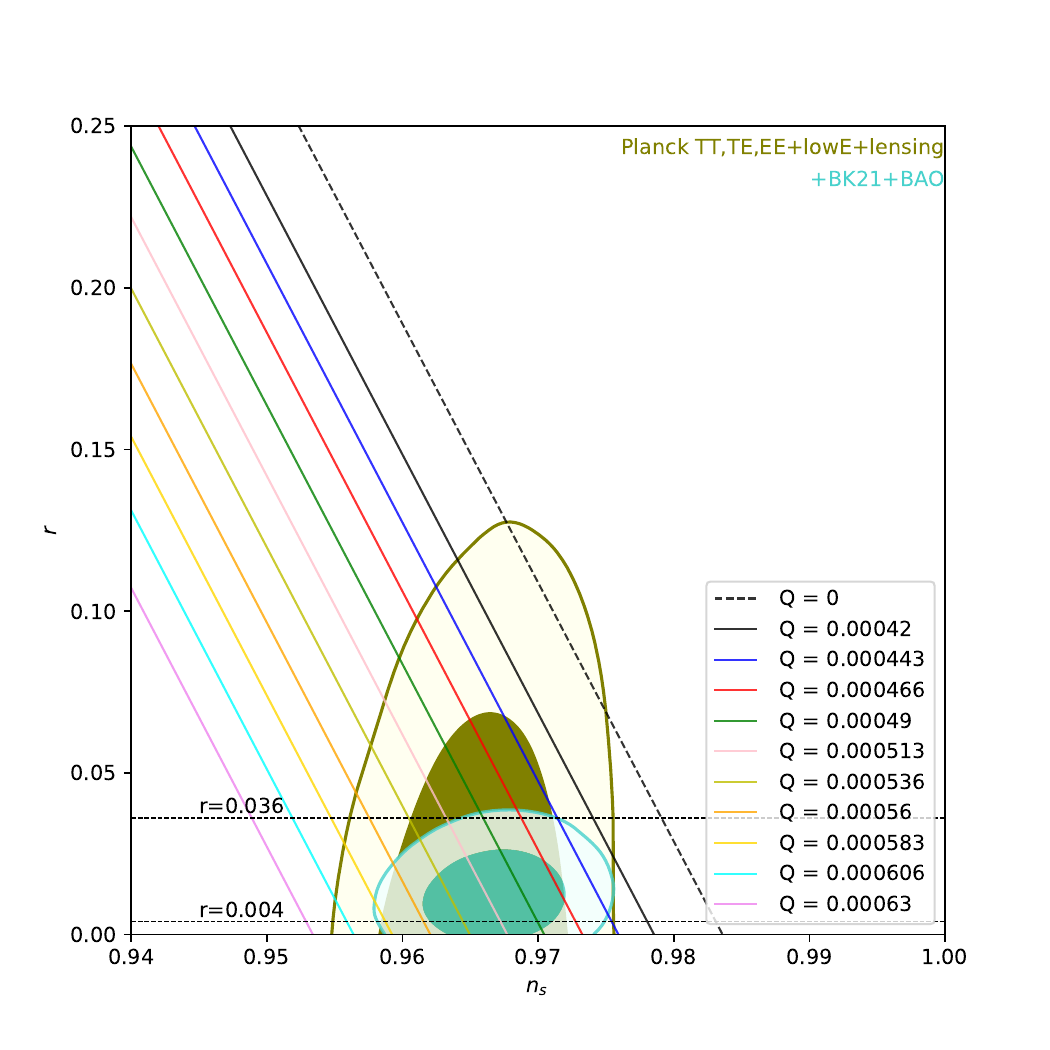}
    \caption{Quartic}
     \end{subfigure}}
\caption{\label{rnswli} Effect of dissipation on the tensor to scalar ratio $(r)$ of quadratic and quartic WLI model for a range of scalar spectral index ($n_s$) with 95$\%$ and 68$\%$ CL of joint Planck TT,TE,EE+lowE+lensing and BK18+BAO.}
\end{figure}

WLI model can overcome the issues of both dissipation and thermal correction by directly coupling the inflaton to a few lighter fields. Here, a Pseudo-Nambu-Goldstone-Boson (PNGB) of a broken U(1) gauge symmetry is considered as a candidate for inflaton. The mass of such a boson produced through a process similar to Little Higgs electro-weak symmetry breaking \cite{eft,wli1} is safeguarded from higher order radiative corrections. This model also has the potential to describe the dark matter production in the universe \cite{eft,wlidm}. 
We consider the simplest case where two complex Higgs fields $\phi_1$ and $\phi_2$ such that \cite{wli1}
\begin{eqnarray}
<\phi_1> \ = \ <\phi_2> \ =  \ \frac{M}{\sqrt{2}},\\
\phi_1 = \frac{M}{\sqrt{2}} e^{i \frac{\phi}{M}} \  \mbox{and} \  \phi_2 = \frac{M}{\sqrt{2}} e^{-i \frac{\phi}{M}}.
\end{eqnarray}
Here $\phi$ is the relative phase of the fields $\phi_1$ and $\phi_2$ and is considered as inflaton (for details see \cite{eft,wli1}). The interaction between fermions ($\psi_{1,2}$) and the Higgs fields can be described by assuming identical coupling and interchange symmetry ($\phi_1 \leftrightarrow i\phi_2$ and $\psi_1{_{L,R}} \leftrightarrow \psi_2{_{L,R}}$) as,
 \begin{eqnarray}
 \mathcal{L}_{int} &=& \mathcal{L}_{\phi \psi}  + \mathcal{L}_{\psi \sigma}, \\
- \mathcal{L}_{\phi \psi} &=& \frac{g}{\sqrt{2}} (\phi_1 + \phi_2)\bar \psi_{1L}\psi_{1R}  + \frac{ig}{\sqrt{2}} (\phi_1 - \phi_2)\bar \psi_{2L}\psi_{2R},\\
 \mathcal{L}_{\psi \sigma} &=& - h\sigma \sum_{i=1,2} (\bar \psi_{iL} \psi_{\sigma R} + \bar \psi_{\sigma L} \psi_{i R}).
\end{eqnarray}
There are no other non-renormalizable terms present in the Lagrangian which is an added bonus to this model. Also, the thermal corrections and the local self energy of the inflaton cancel each other. However, the nonlocal effects like retarded self energy ($\Sigma_R$) of inflaton have a non negligible contribution to the dissipation \cite{wli1},
\begin{eqnarray}
\gamma(T) &=& \int d^4x' \Sigma_R(x,x')(t'-t).
\end{eqnarray}
The major contribution to the dissipation comes from the on-shell decay of fermions, with the dissipation coefficient having a linear dependence on temperature \cite{wli1} i.e.,
\begin{eqnarray}
\gamma(T) &=& c_T T
\end{eqnarray}
where, $c_T \simeq \frac{\alpha(h)g^2}{h^2}$ and $\alpha(h) = \frac{3}{1-0.34 \log(h)}$. We study the quadratic and quartic model of inflation in the warm little inflationary setting for $N=60$ and the results are presented in Table(\ref{wlislow}) and Figure(\ref{rnswli}). Since, WLI lies in the weak dissipation regime ($Q<<1$), $G_{l=1}(Q)\simeq1$.
\begin{table*}
\centering
\caption{\label{wlislow}
Cosmological inflationary parameters of quadratic and quartic warm inflation for WLI in terms of the e-folding number ($N$) and $Q$.}
\begin{tabular*}{\textwidth}{@{}l*{15}{@{\extracolsep{0pt plus10pt}}l}}
\hline
Inflationary parameter & $V(\phi)=\frac{1}{2}m^2\phi^2$ & $V(\phi)=\frac{\lambda}{4}\phi^4$ \\
\hline
$\epsilon_H$  & $ \frac{1}{2N(1+Q)}$ & $\frac{1}{N(1+Q)}$ \\
$\eta_H$ & $\frac{Q}{2N}+Q$ & $\frac{3Q}{2N}+Q$  \\
$n_s^{WI}$ & $1- \frac{3}{N}+\frac{\pi c Q^\frac{5}{4}}{2}\bigg(\frac{13}{2N}+9\bigg)$ & $1- \frac{3}{N}+\frac{\pi c Q^\frac{5}{4}}{2}\bigg(\frac{12}{2N}+9\bigg)$ \\
$r^{WI}$  & $\frac{8}{ NcQ^\frac{1}{4}G_{l=1}(Q)}  $ & $\frac{8}{ NcQ^\frac{1}{4}G_{l=1}(Q)}$\\
\hline
\end{tabular*}
\end{table*}
From the analysis, we show that the newly obtained scalar spectral index and tensor to scalar ratio are consistent with the joint analysis of Planck TT,TE,EE+lowE+lensing and BK18+BAO. Further, we obtain a bound $0.0015<Q<0.0019$ for quadratic model and $0.00046<Q< 0.00056$ for quartic model from the 68\% CL of the joint analysis of Planck18 and BK18. From the analysis, it is evident that the forthcoming CMB estimates also strongly agree with these bounds. Since the obtained $Q<<1$, this model falls in the weak dissipative regime. 

\subsection{Minimal warm inflation (MWI)}
\begin{figure}[h!]
\centering
\makebox[\linewidth]
{
   \begin{subfigure}{0.5\linewidth}
    \includegraphics[width=\linewidth]{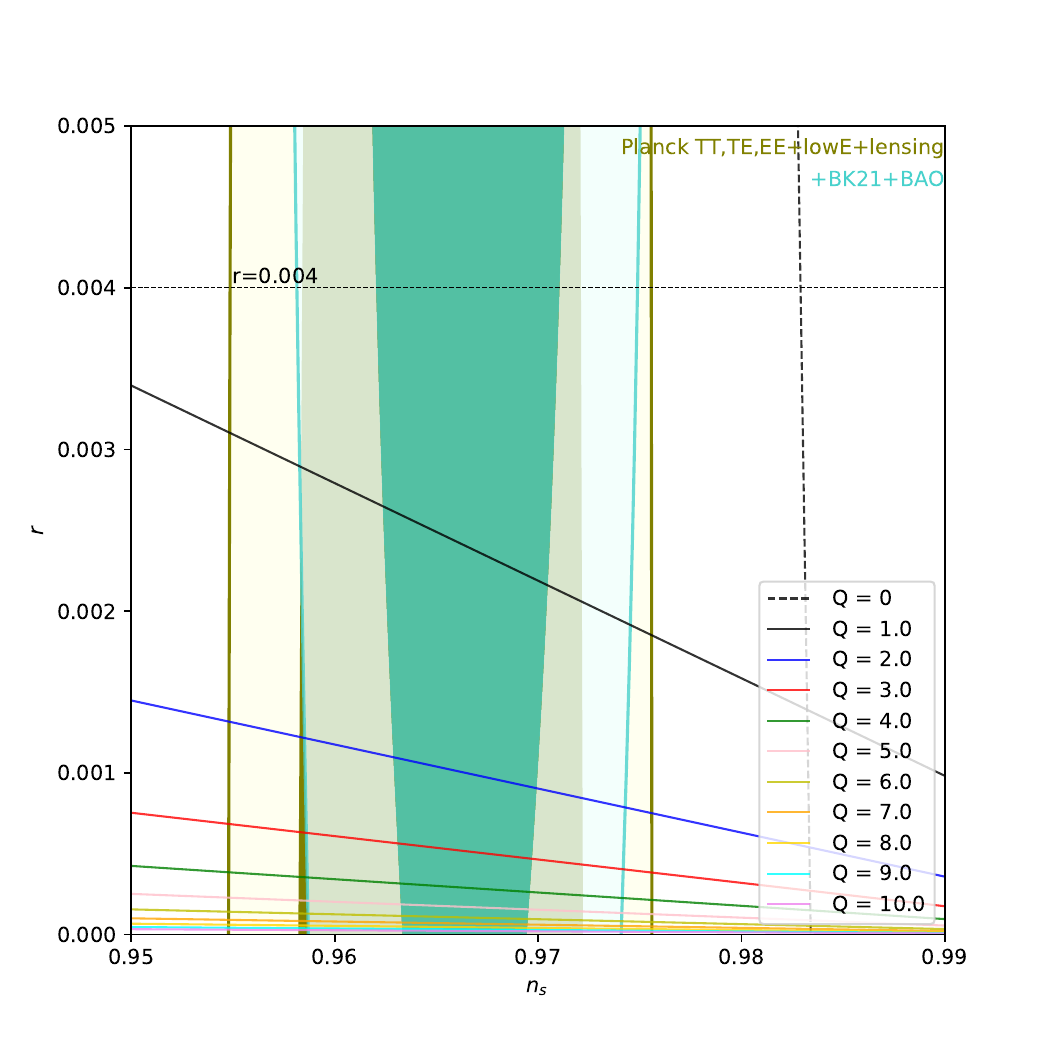}
     \caption{Quadratic}
     \end{subfigure}
     \begin{subfigure}{0.5\linewidth}
    \includegraphics[width=\linewidth]{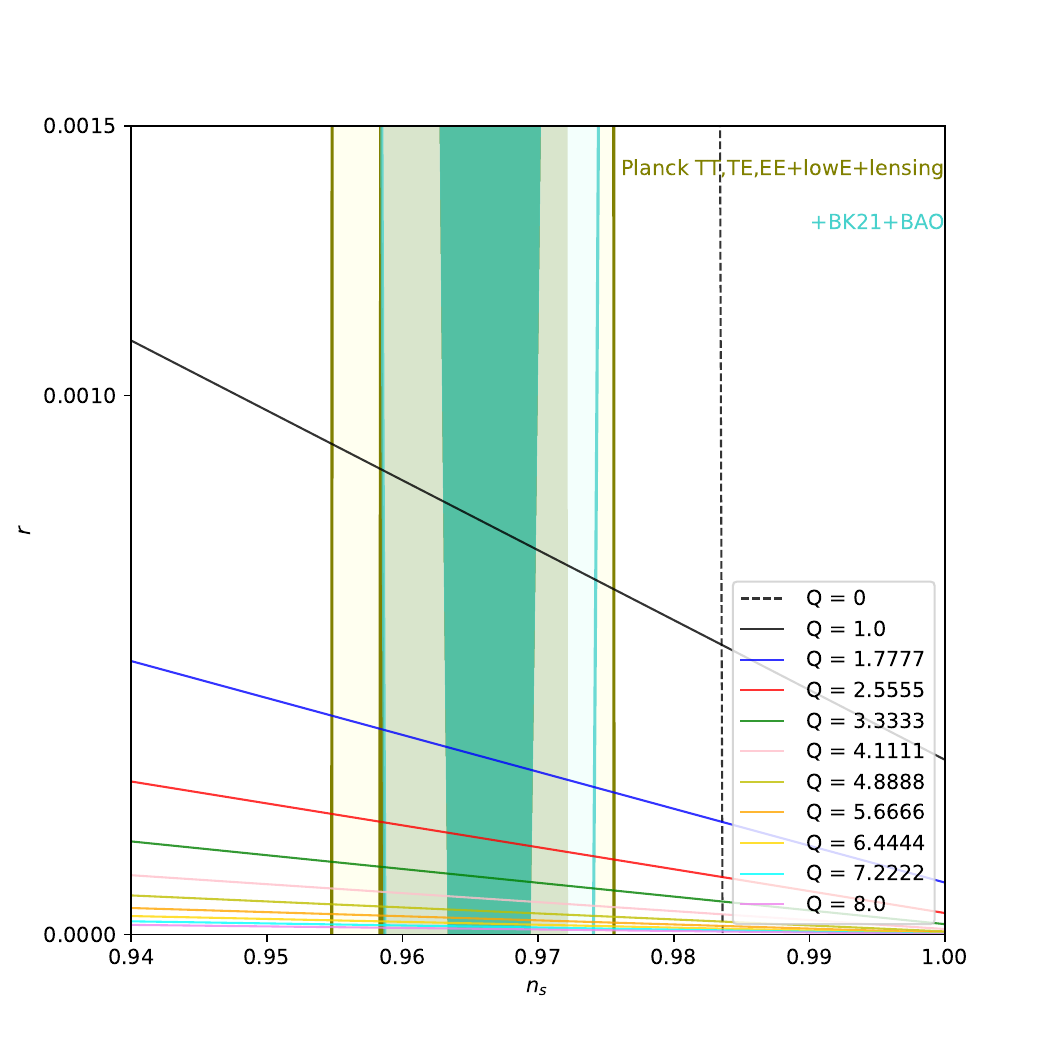}
    \caption{Quartic}
     \end{subfigure}
     
}
\caption{\label{rnsax} Effect of dissipation on the tensor to scalar ratio $(r)$ of quadratic and quartic MWI model for a range of scalar spectral index ($n_s$) with 95$\%$ and 68$\%$ CL of joint Planck TT,TE,EE+lowE+lensing and BK18+BAO.}
\end{figure}

Currently, intensive research is carried out in the direction of a model with minimum ingredients where axion is considered as inflaton and is coupled to non-Abelian gauge fields, for example, Yang-Mills (YM) field. The gauge vacua can mediate sphaleron transitions leading to friction and hence dissipation \cite{kim,sph}. Dissipative axion in such a minimalistic model can effortlessly arise from string theory where its shift symmetry is known to prevent any backreaction and related thermal mass corrections. These interesting features give an upper hand to this model compared to other models proposed so far.  The interaction between axion ($\Phi$) and an arbitrary Yang-Mills group ($G^{\mu \nu}_a$) is best described by the Lagrangian \cite{kim},
\begin{equation}
\mathcal{L}_{int} = \frac{\alpha}{16 \pi} \frac{\Phi}{f} \tilde G^{\mu \nu}_a G^a_{\mu \nu} \label{lgm}
\end{equation}
where $f$ is the axion decay constant and $\alpha$ is related to the Yang-Mills gauge coupling $g$ as $\alpha = \frac{g^2}{4\pi}$. The resulting interaction has a dissipation coefficient \cite{kim,sph,chetan}
\begin{eqnarray}
\gamma(T)=\kappa(\alpha,N_c,N_f)\frac{\alpha^5T^3}{f^2},
\end{eqnarray}
where $\kappa$ is a dimensionless quantity that depends on the number of colours ($N_c$) and flavours ($N_f$) of the gauge group. To check the feasibility of the interaction in (\ref{lgm}), we study this model with quadratic and quartic inflaton potential for $N=60$ and $G(Q)=G_{l=3}(Q)$. The results are presented in Table(\ref{axslow}) and Figure(\ref{rnsax}), where $G'_{l=3}(Q) = \frac{\mathrm{d}G_{l=3}(Q)}{\mathrm{d}Q} = 9.6930  Q^{0.946} + 0.5506 Q^{3.336}$.
\begin{table}
\caption{\label{axslow}
Cosmological inflationary parameters of quadratic and quartic warm inflation for MWI in terms of the e-folding number ($N$) and $Q$.}
\begin{tabular*}{\textwidth}{@{}l*{15}{@{\extracolsep{0pt plus10pt}}l}}
\hline
Inflationary  &  $V(\phi)=\frac{1}{2}m^2\phi^2$ & $V(\phi)=\frac{\lambda}{4}\phi^4$ \\
parameter& & \\
\hline
$\epsilon_H$  & $\frac{1}{2NQ}$ & $\frac{1}{NQ}$ \\ \\
$\eta_H$ & $\frac{1}{2NQ}+3$ & $\frac{3}{2NQ}$  \\ \\
$n_s^{WI}$ & $1- \frac{3}{8QN} $ & $1- \frac{9}{14(1+Q)N}$ \\ \\
&$ + \frac{2QG'_{l=3}(Q)}{7NG_{l=3}(Q)}$ & $ + \frac{QG'_{l=3}(Q)}{7NG_{l=3}(Q)} $. \\ \\
$r^{WI}$  & $\frac{8}{ Nc \sqrt{3\pi}Q^\frac{1}{4}G_{l=3}(Q)}$ & $\frac{8}{ Nc \sqrt{3\pi}Q^\frac{1}{4}G_{l=3}(Q)}$\\
\hline
\end{tabular*}
\end{table}
These results are consistent with joint Planck18 and BICEP/Keck analysis \cite{10} and therefore this model is viable for warm inflation. We derive a range $1<Q< 8$ for quadratic model and $1<Q<10$ for quartic model from the 68\% CL of joint Planck 18 and BK18. This is the highest value of $Q$ obtained for warm inflation in the strong dissipative regime. Apart from this, the upcoming CMB estimates from LiteBird and CMB S4 ($r \leq 0.004$) also supports MWI in the strong dissipative regime.

\section{Effect on Hubble parameter}
\begin{figure}[h!]
\centering
\makebox[\linewidth]
{
   \begin{subfigure}{0.5\linewidth}
    \includegraphics[width=\linewidth]{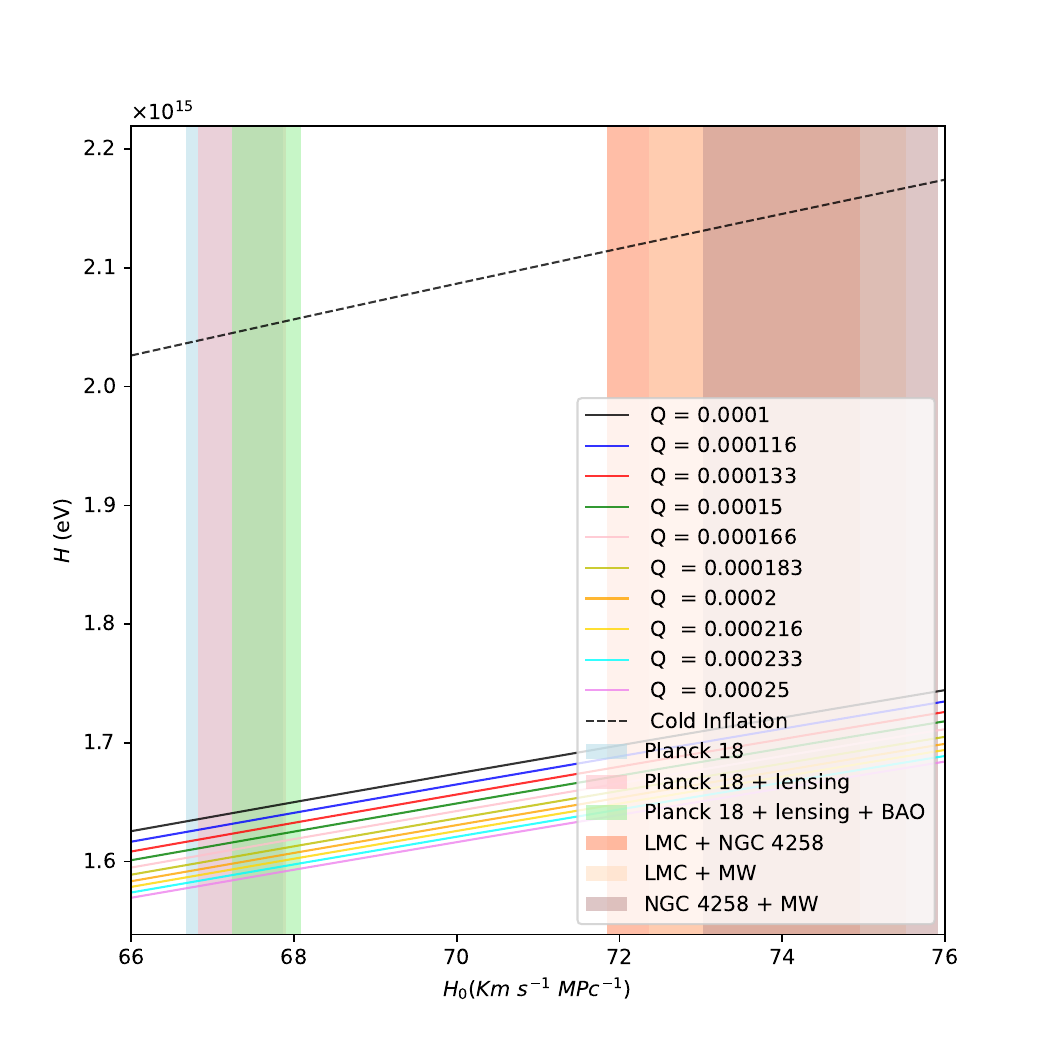}
     \caption{Quadratic}
     \end{subfigure}
     \begin{subfigure}{0.5\linewidth}
    \includegraphics[width=\linewidth]{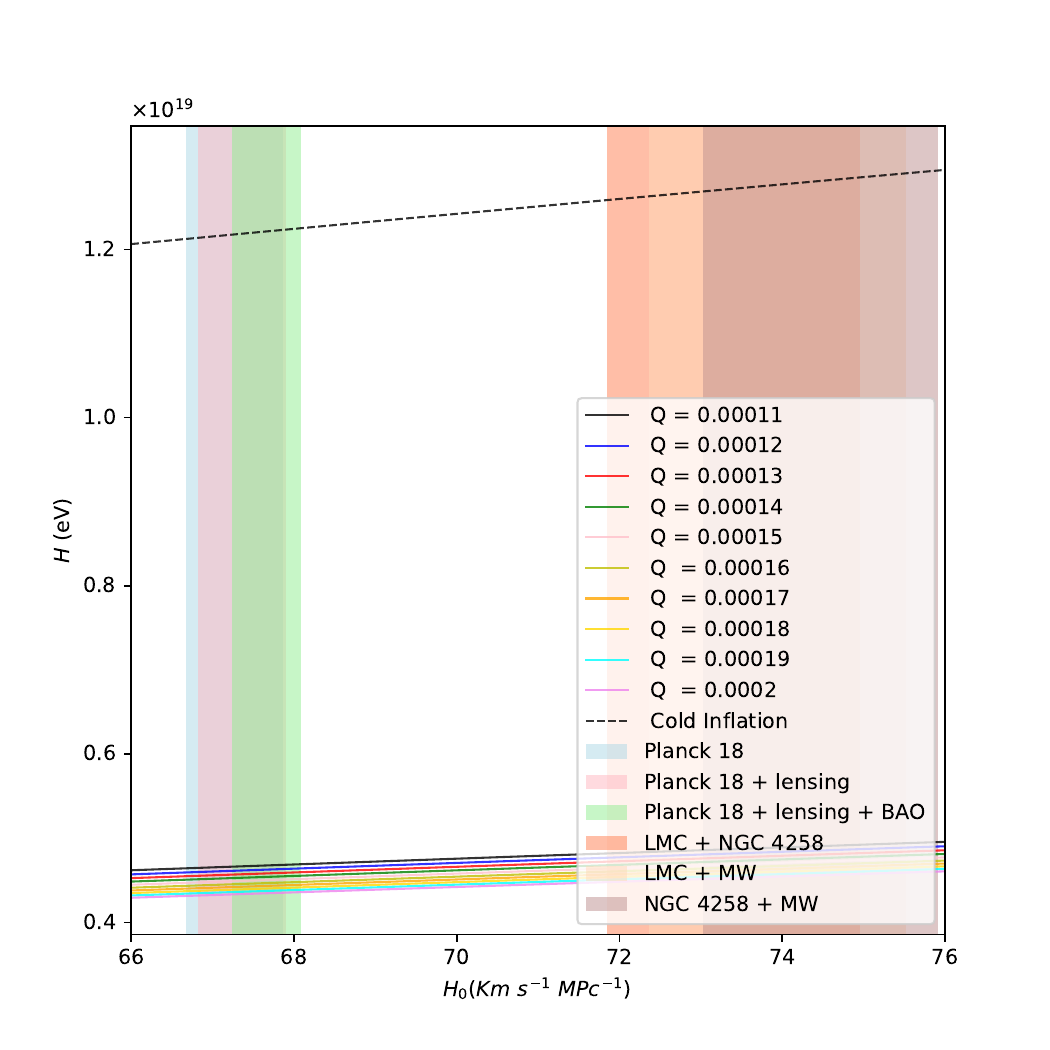}
    \caption{Quartic}
     \end{subfigure}
}\par
\makebox[\linewidth]
{
   \begin{subfigure}{0.5\linewidth}
    \includegraphics[width=\linewidth]{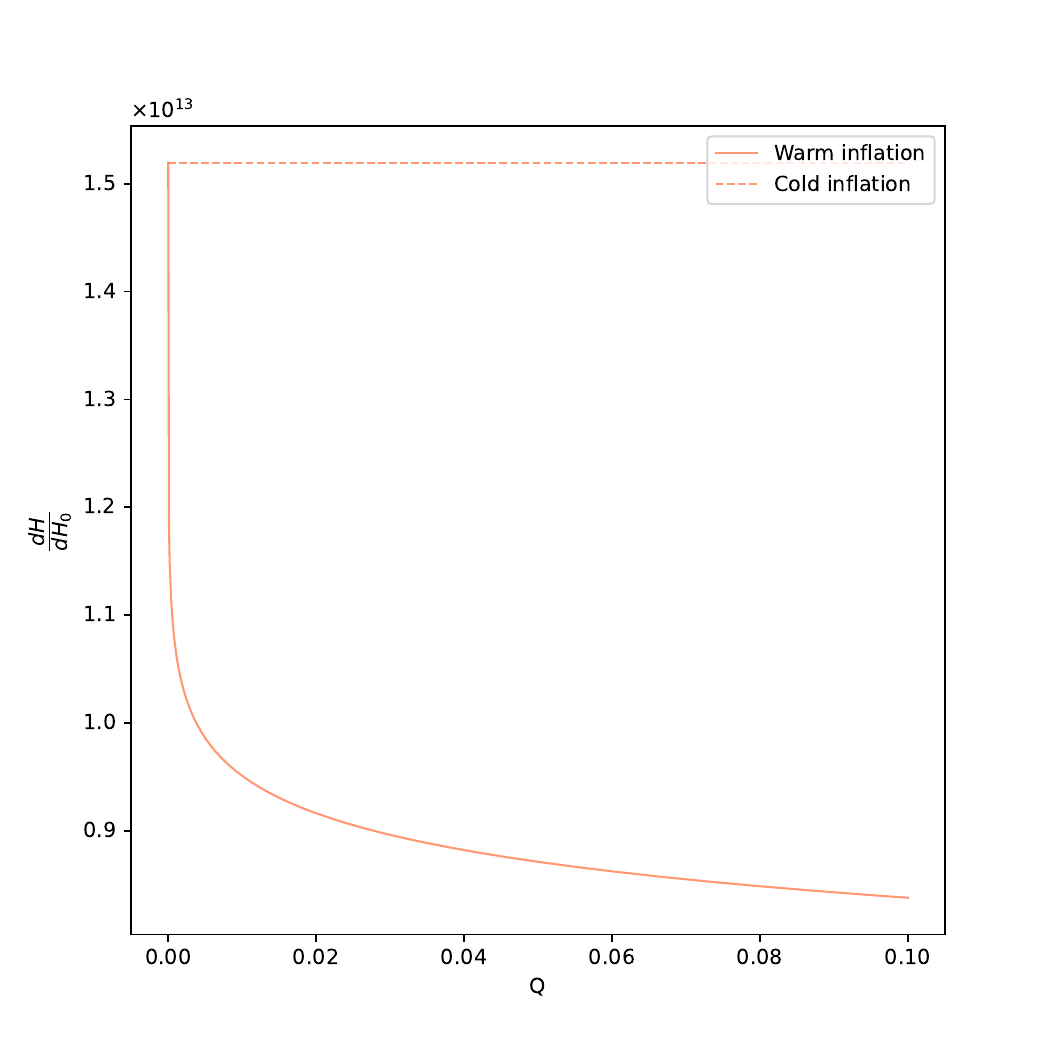}
     \caption{Quadratic}
     \end{subfigure}
     \begin{subfigure}{0.5\linewidth}
    \includegraphics[width=\linewidth]{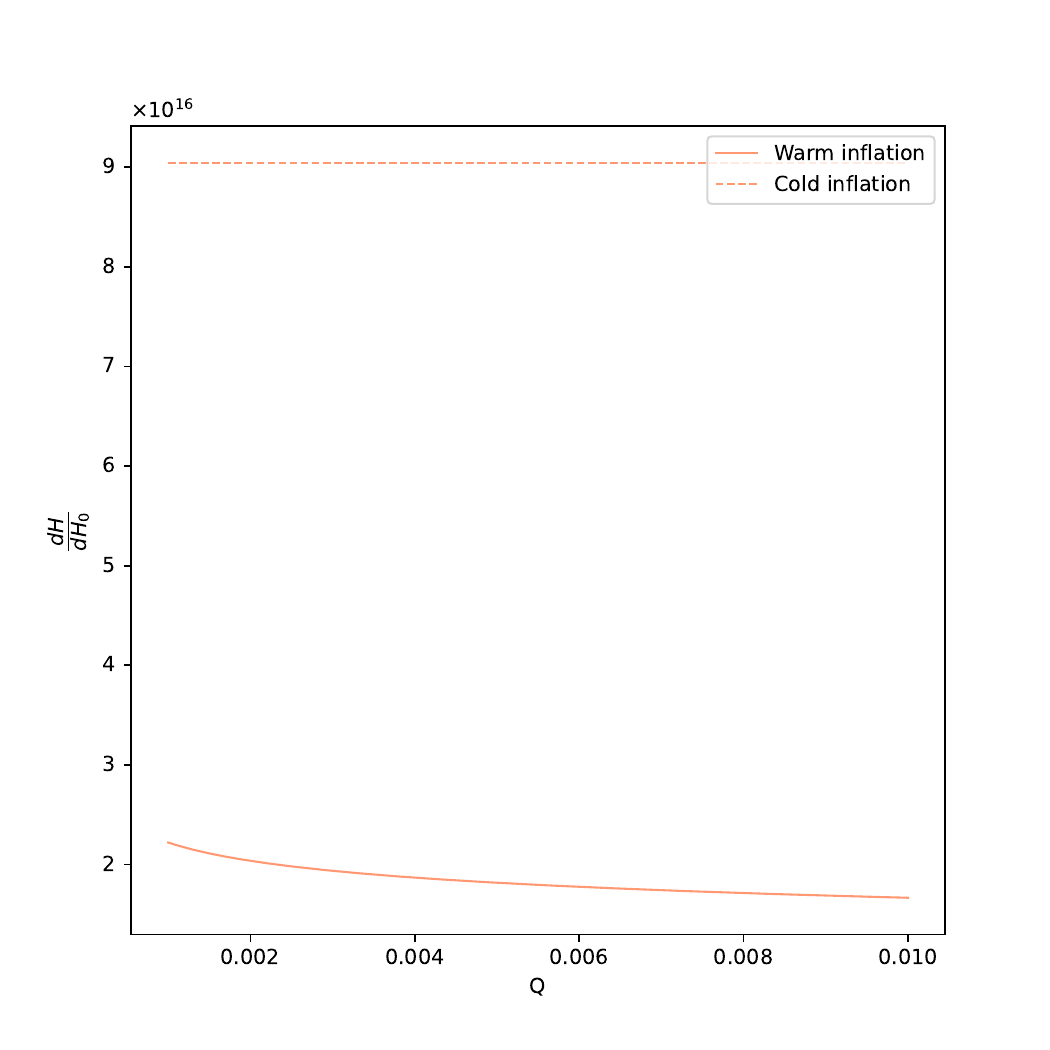}
    \caption{Quartic}
     \end{subfigure}
}
\caption{\label{HTtwo} Dissipation effect on the Hubble parameter during inflation (H) for a range of present Hubble parameter ($H_0$) for quadratic (a) and quartic (b) in TSM model with Planck 2018 and SH0ES data (top panel).  The effect of Q on the slope ($s = \frac{dH}{dH_0}$ ) for quadratic (c) and quartic (d) in TSM model with Planck 2018 and SH0ES data (bottom panel).}
\end{figure}

\begin{figure}[h!]
\centering
\makebox[\linewidth]
{
   \begin{subfigure}{0.5\linewidth}
    \includegraphics[width=\linewidth]{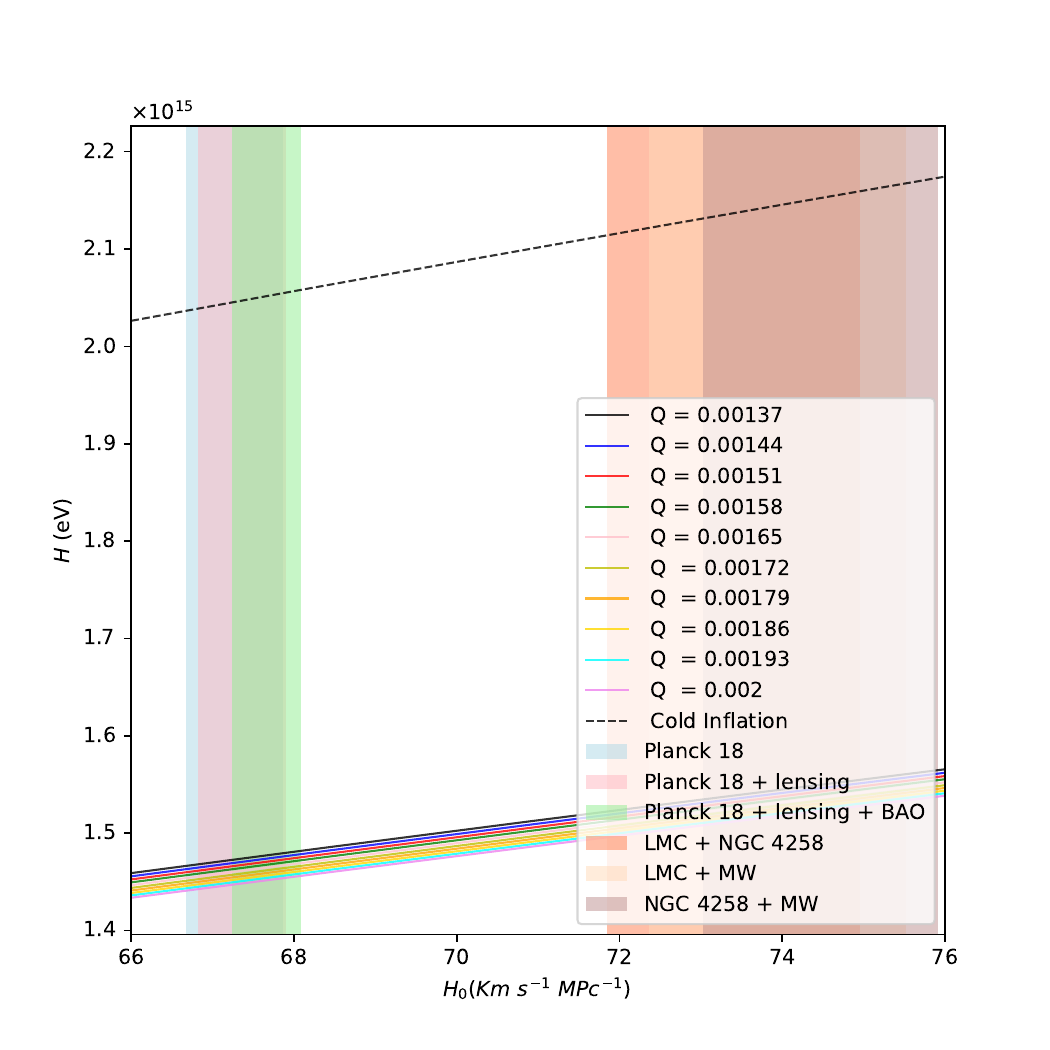}
     \caption{Quadratic}
     \end{subfigure}
     \begin{subfigure}{0.5\linewidth}
    \includegraphics[width=\linewidth]{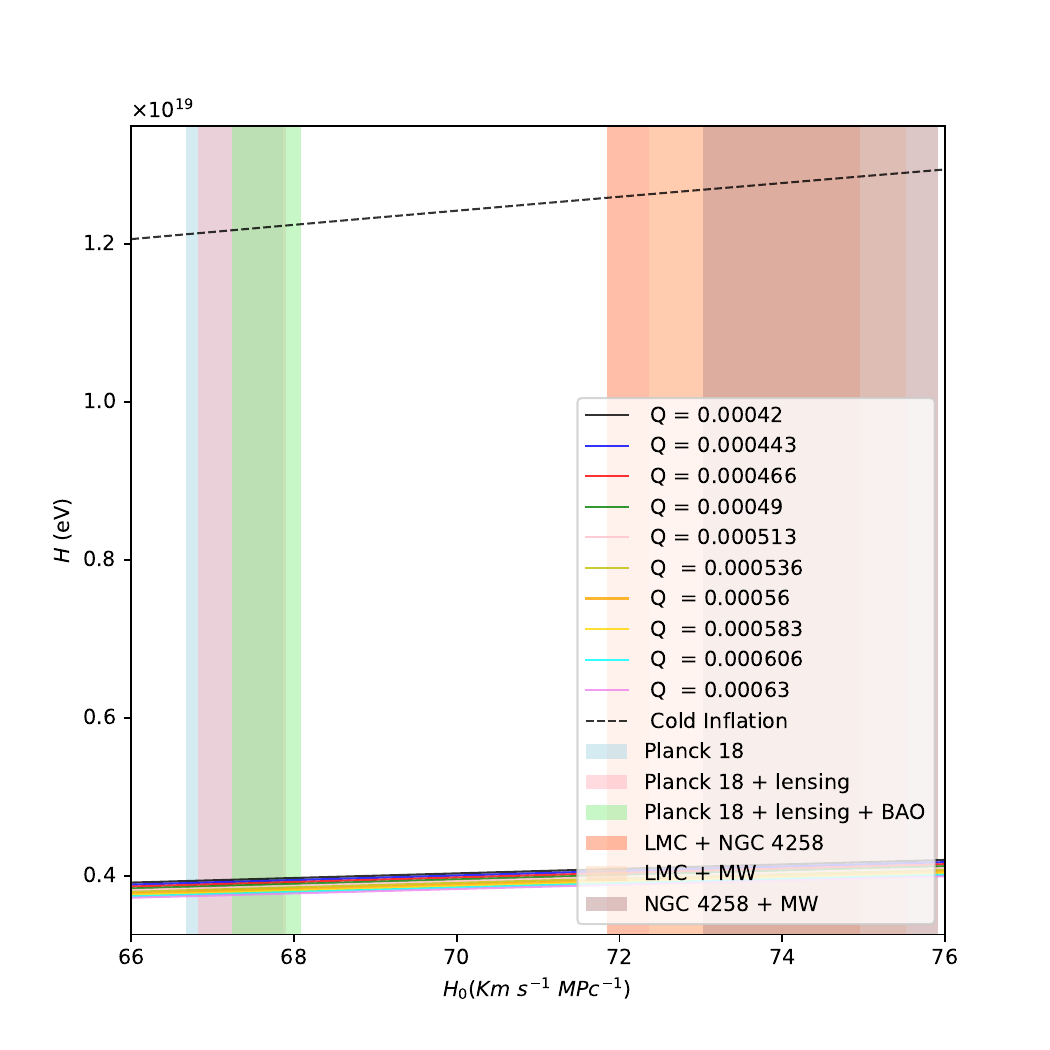}
    \caption{Quartic}
     \end{subfigure}
}\par
\makebox[\linewidth]
{
   \begin{subfigure}{0.5\linewidth}
    \includegraphics[width=\linewidth]{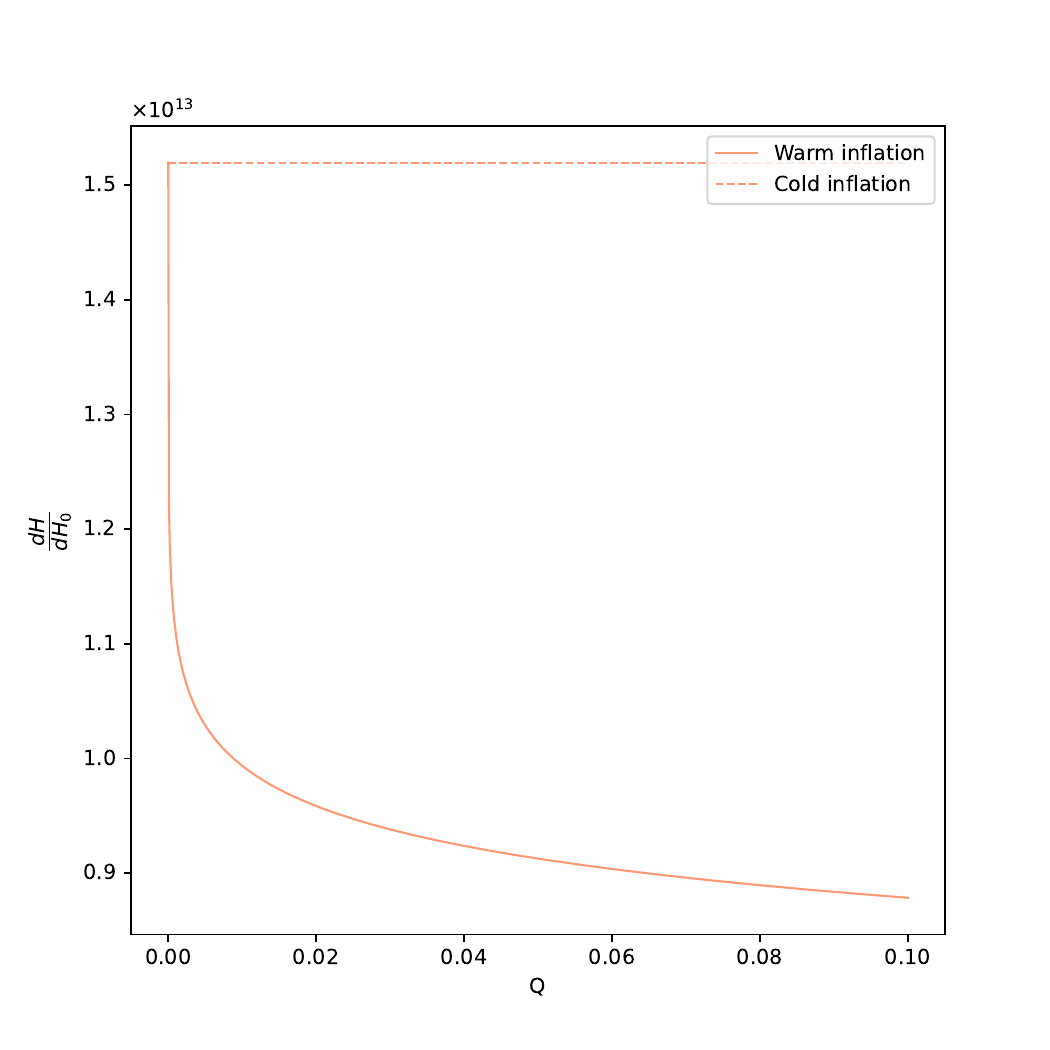}
     \caption{Quadratic}
     \end{subfigure}
     \begin{subfigure}{0.5\linewidth}
    \includegraphics[width=\linewidth]{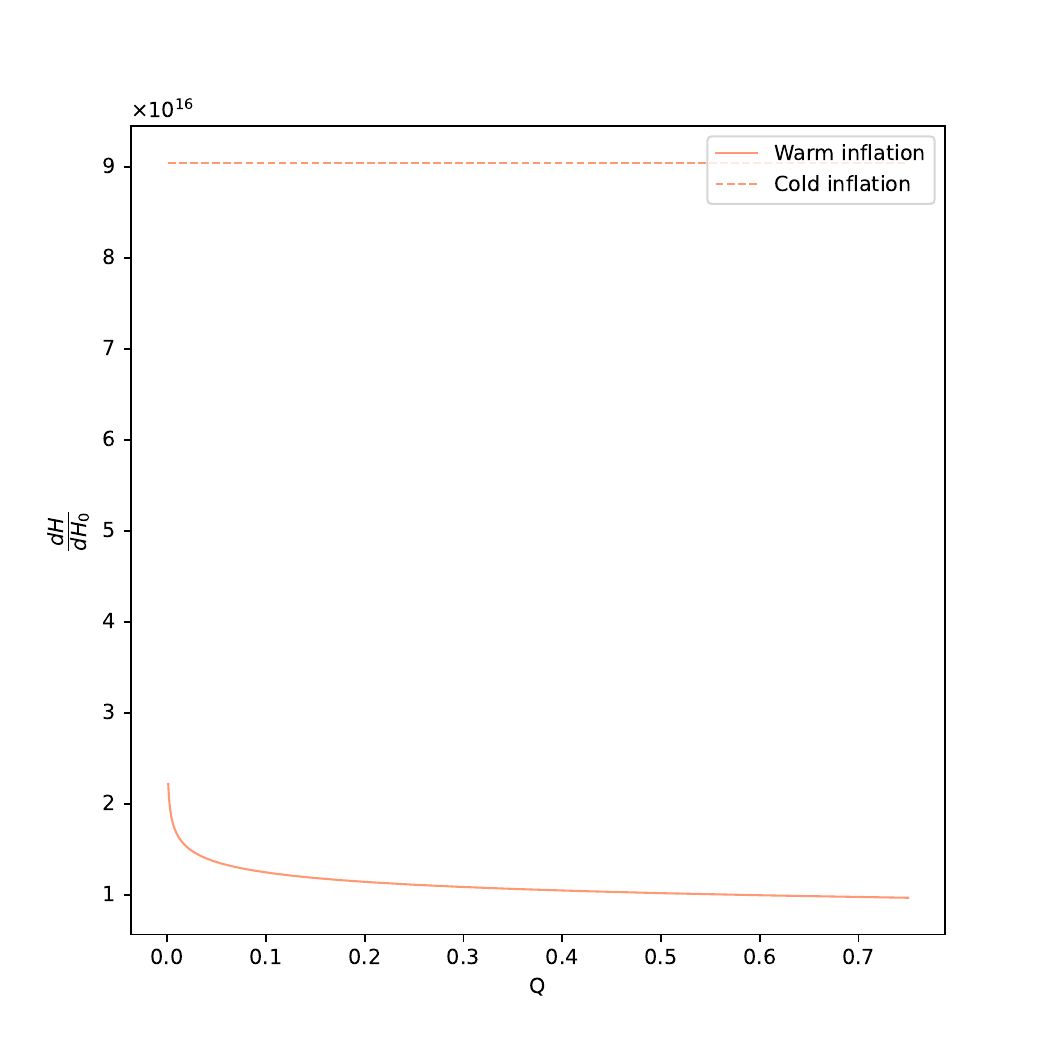}
    \caption{Quartic}
     \end{subfigure}
}
\caption{\label{HTwli} Dissipation effect on the Hubble parameter during inflation (H) for a range of present Hubble parameter ($H_0$) for quadratic (a) and quartic (b) in WLI model with Planck 2018 and SH0ES data (top panel).  The effect of Q on the slope ($s = \frac{dH}{dH_0}$ ) for quadratic (c) and quartic (d) in WLI model with Planck 2018 and SH0ES data (bottom panel).}
\end{figure}

\begin{figure}[h!]
\centering
\makebox[\linewidth]
{
   \begin{subfigure}{0.5\linewidth}
    \includegraphics[width=\linewidth]{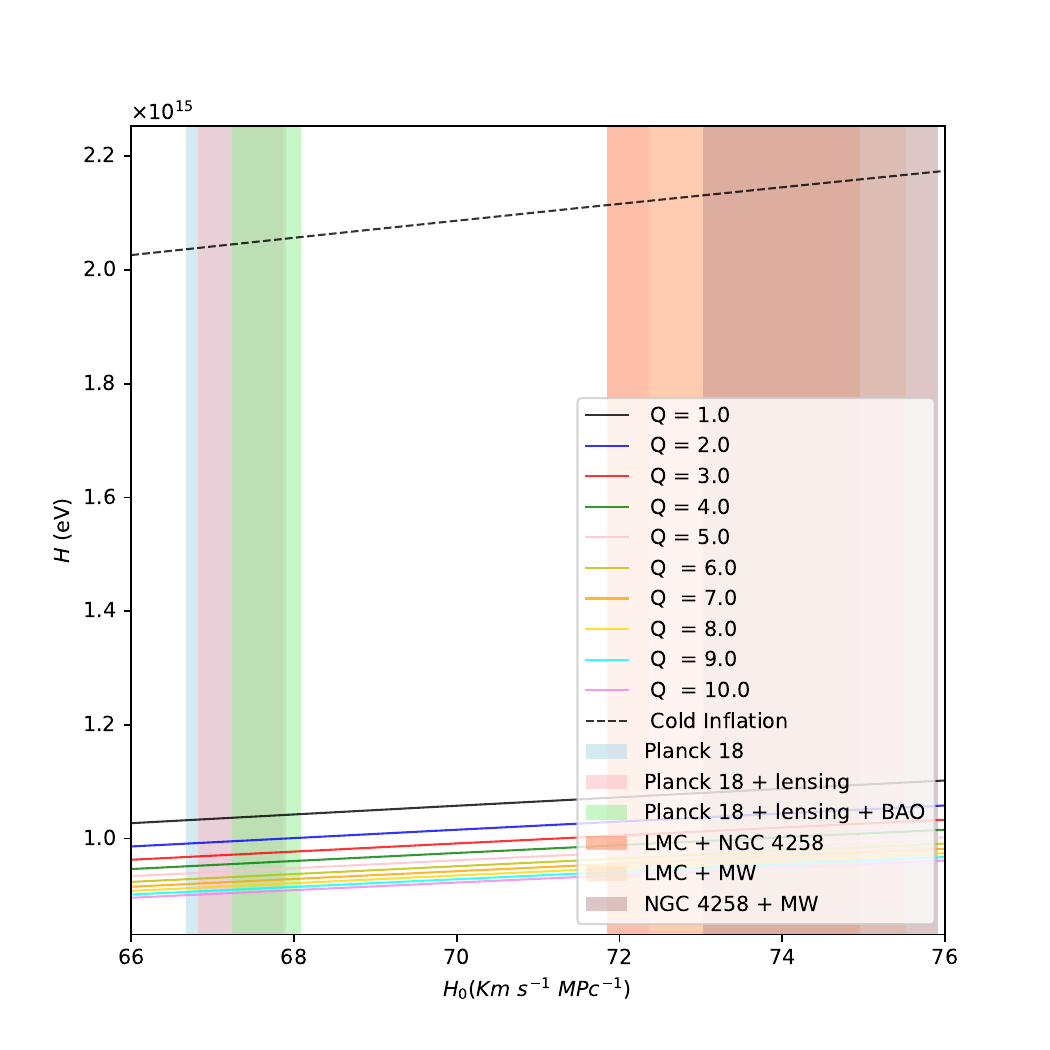}
     \caption{Quadratic}
     \end{subfigure}
     \begin{subfigure}{0.5\linewidth}
    \includegraphics[width=\linewidth]{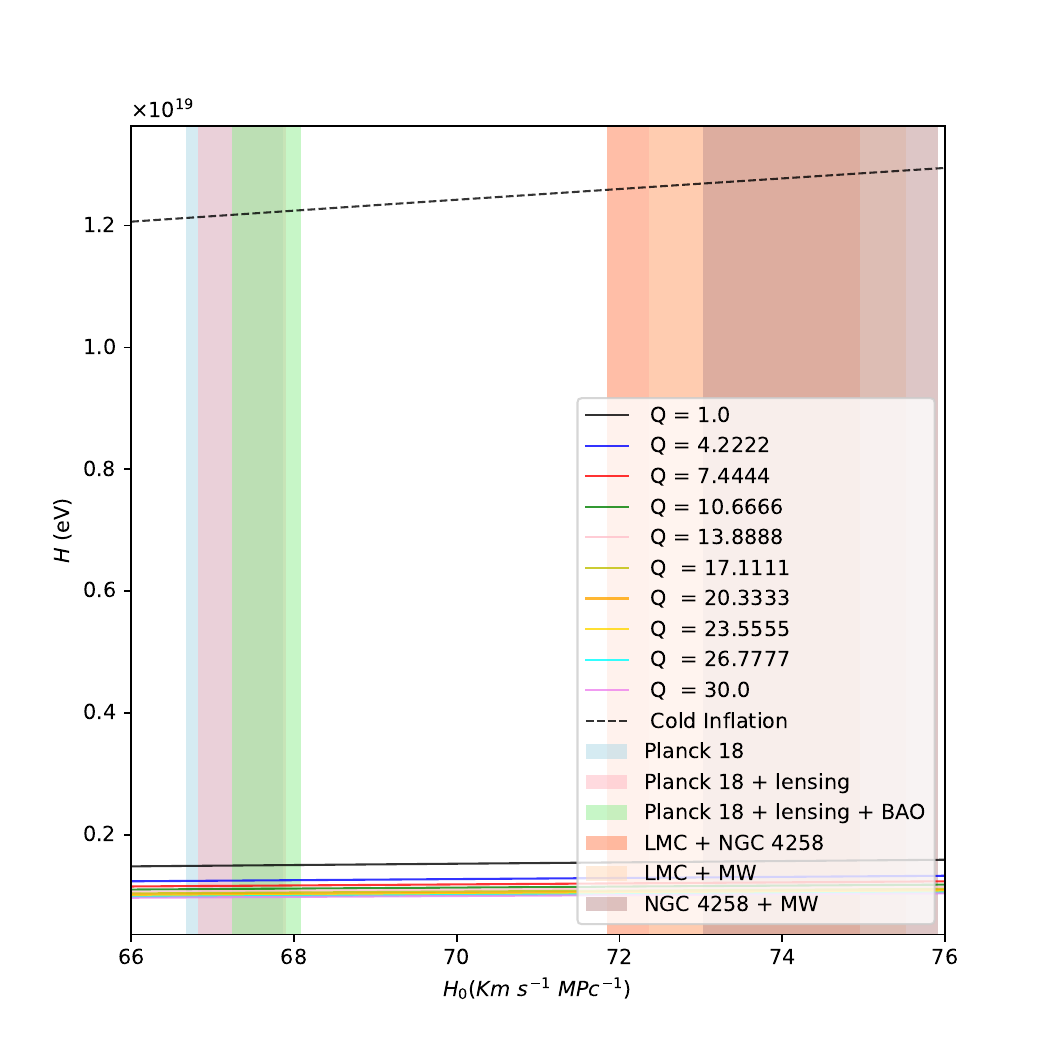}
    \caption{Quartic}
     \end{subfigure}
}\par
\makebox[\linewidth]
{
   \begin{subfigure}{0.5\linewidth}
    \includegraphics[width=\linewidth]{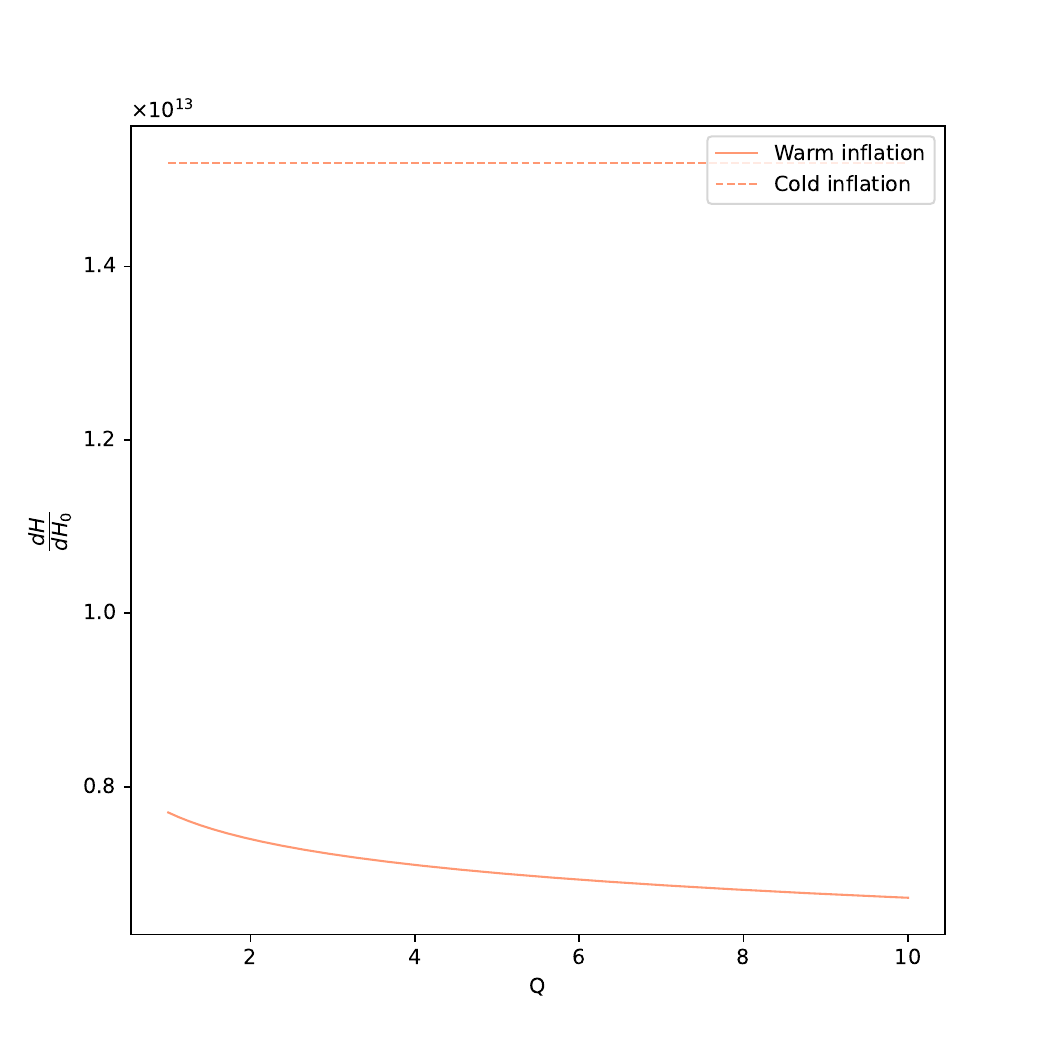}
     \caption{Quadratic}
     \end{subfigure}
     \begin{subfigure}{0.5\linewidth}
    \includegraphics[width=\linewidth]{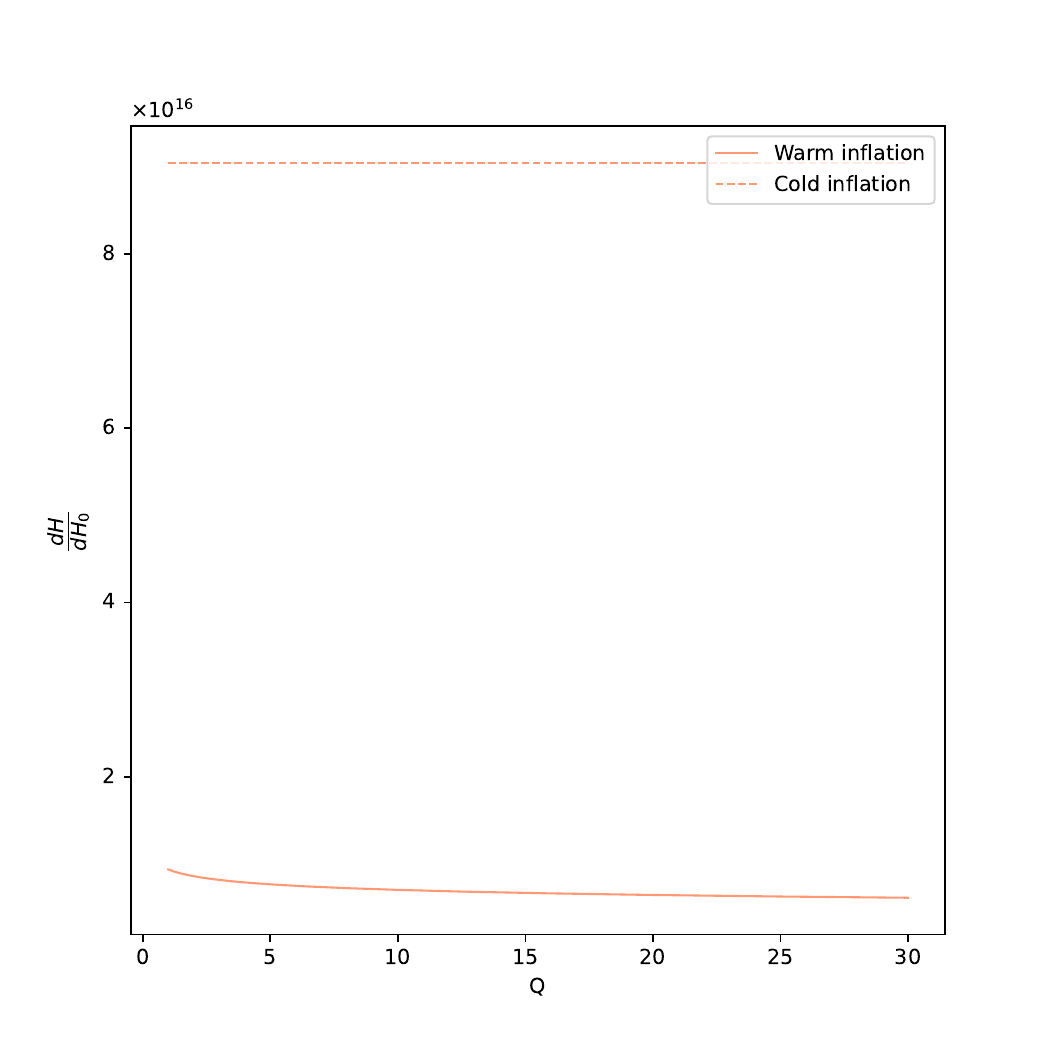}
    \caption{Quartic}
     \end{subfigure}
}
\caption{\label{HTax} Dissipation effect on the Hubble parameter during inflation (H) for a range of present Hubble parameter ($H_0$) for quadratic (a) and quartic (b) in MWI model with Planck 2018 and SH0ES data (top panel).  The effect of Q on the slope ($s = \frac{dH}{dH_0}$ ) for quadratic (c) and quartic (d) in MWI model with Planck 2018 and SH0ES data (bottom panel).}
\end{figure}

The mismatch between the $\Lambda$CDM model based indirect measurement of present Hubble parameter and the supernova based direct measurement of present Hubble parameter results in a major tension in cosmology.  The former infers a lower value ($H_0$ = 67.36 $\pm$ 0.54 km s$^{-1}$ Mpc$^{-1}$) \cite{planck} and the latter reports a higher value ($H_0$=74.03 $\pm$ 1.04 km s$^{-1}$ Mpc$^{-1}$) \cite{reiss} known as Hubble tension. The conventional cold inflationary scenario is not inherent in standard model of cosmology, rather it is added as an extension to address some of the major problems in standard cosmology. There are several challenges to $\Lambda$CDM model and many alternatives have been proposed to address this issue \cite{chl, model, odi}. Hence, CI may not be a right choice of model to explain this discrepancy in the present value of the Hubble parameter. Various measurements and predictions of the present Hubble parameter with modified theories and improved observations cluster around the higher ($H_0 \sim$ 74 km s$^{-1}$ Mpc$^{-1}$) and lower values ($H_0 \sim$ 68 km s$^{-1}$ Mpc$^{-1}$) \cite{dv1,dv2}. Other approaches to resolve Hubble tension did not succeed completely and lacks satisfactory conceptual confrontation for the existence of two different values for $H_0$ \cite{kami}. Therefore, the research community has not reached a consensus on this discrepancy \cite{tale}. The concept of WI is based on the idea of evolution and decay of inflaton in a static radiation bath throughout inflation where thermal fluctuations dominate over the quantum field fluctuations \cite{kamal,story}. The fluctuations take the form,
\begin{eqnarray}\label{pert}
<\delta \phi^2> \  \sim
\cases{
 \frac{H^2}{2 \pi} & for CI \cr
 \frac{3HT}{4\pi} & for WI with $Q<<1$ \cr
\frac{\sqrt{\gamma H} T}{2\pi^2} &  for WI with $Q>>1$. \cr }
\end{eqnarray}
During warm inflation, the universe tries to attain an equilibrium temperature $T > H$ which can be maintained only for a short duration as the expansion can cause a dilution of the decay products in the thermal bath \cite{SKS}. These interactions and the resultant thermal fluctuations in the early universe can affect the evolution of the Hubble parameter. Interestingly, this motivates to study the Hubble parameter using the supersymmetric and string based warm inflationary models and examine its role in alleviating the Hubble tension.\\ 
\\
We obtain a relation between the Hubble parameter during warm inflation and the present Hubble parameter for quadratic and quartic inflaton potential \cite{anu,bour} (for a detailed explanation see appendix),
\begin{eqnarray} \label{sim}
H = \cases{
          \bigg( \frac{12 \pi \Omega_\Lambda}{F(Q)}\bigg)^\frac{1}{4} \sqrt{m_{pl}H_0}, & for  $V(\phi)=\frac{1}{2}m^2\phi^2$ \cr
          \bigg( \frac{48 \Omega_{\Lambda}}{\lambda F(Q) }\bigg)^\frac{1}{4} \sqrt{\pi m_{pl} H_0}, & for $ V(\phi)=\frac{ \lambda}{4}\phi^4.$ \cr }
\end{eqnarray}
\begin{table}[h]
\begin{center}
\caption{\label{HTFQ}
Effect of dissipation on Hubble parameter for various warm inflationary models}
\begin{tabular*}{0.55\textwidth}{@{}l*{15}{@{\extracolsep{0pt plus10pt}}l}}
\hline
Model  & \multicolumn{2}{c}{$F(Q)$} \\
& $V(\phi)=\frac{1}{2}m^2\phi^2$ & $V(\phi)=\frac{\lambda}{4}\phi^4$ \\
\hline
DMM  & $1+\frac{\sqrt{3}cQ^\frac{1}{4}}{\pi}$ & $1+\frac{3cQ^{\frac{3}{4}}}{\pi^2}$ \\
TSM & $1+\frac{3cQ^\frac{1}{4}}{2}$ & $1+\frac{9c^2Q^\frac{1}{2}}{4}$  \\
WLI & $1+\frac{3cQ^\frac{1}{4}}{2}$ &$1+\frac{9c^2Q^\frac{1}{2}}{4}$ \\
MWI & $1+\frac{\sqrt{3}cQ^\frac{1}{4}}{\pi}$ & $1+\frac{3cQ^\frac{3}{4}}{\pi^2}$  \\
\hline
\end{tabular*}
\end{center}
\end{table} 
The obtained result shows that the Hubble parameter during warm inflation ($H$) decreases drastically compared to cold inflation with the increasing strength of dissipation $(Q)$. The presence of $F(Q)$ in (\ref{sim}) implies that the dissipation during warm inflation has a profound impact on the Hubble parameter. $F(Q)$ can take various forms depending upon the nature of interaction and inflaton potential. The different possible $F(Q)$ are given in Table(\ref{HTFQ}).  To differentiate the effect of warm inflation on $H$ from that of cold inflation, we compute the rate of change of $H$ with respect to $H_0$. We refer to this quantity as slope $(s)$ of the Hubble parameter ($H$) and it is given by
\begin{eqnarray}
s&=&\frac{dH}{dH_0} \\ &=&  \cases{
             \frac{1}{2} \bigg(\frac{12 \pi \Omega_\Lambda}{F(Q)}\bigg)^\frac{1}{4} \sqrt{\frac{m_{pl}}{H_0}} & for $V(\phi)=\frac{1}{2}m^2\phi^2$ \cr
           \frac{1}{2} \bigg( \frac{48 \Omega_\Lambda}{\lambda F(Q)} \bigg)^\frac{1}{4} \sqrt{\frac{\pi m_{pl}}{H_0}} & for $V(\phi)=\frac{ \lambda}{4}\phi^4.$ \cr}
\end{eqnarray}\label{s}\\
\begin{figure}[h!]
\centering
\makebox[\linewidth]
{
   \begin{subfigure}{0.5\linewidth}
    \includegraphics[width=\linewidth]{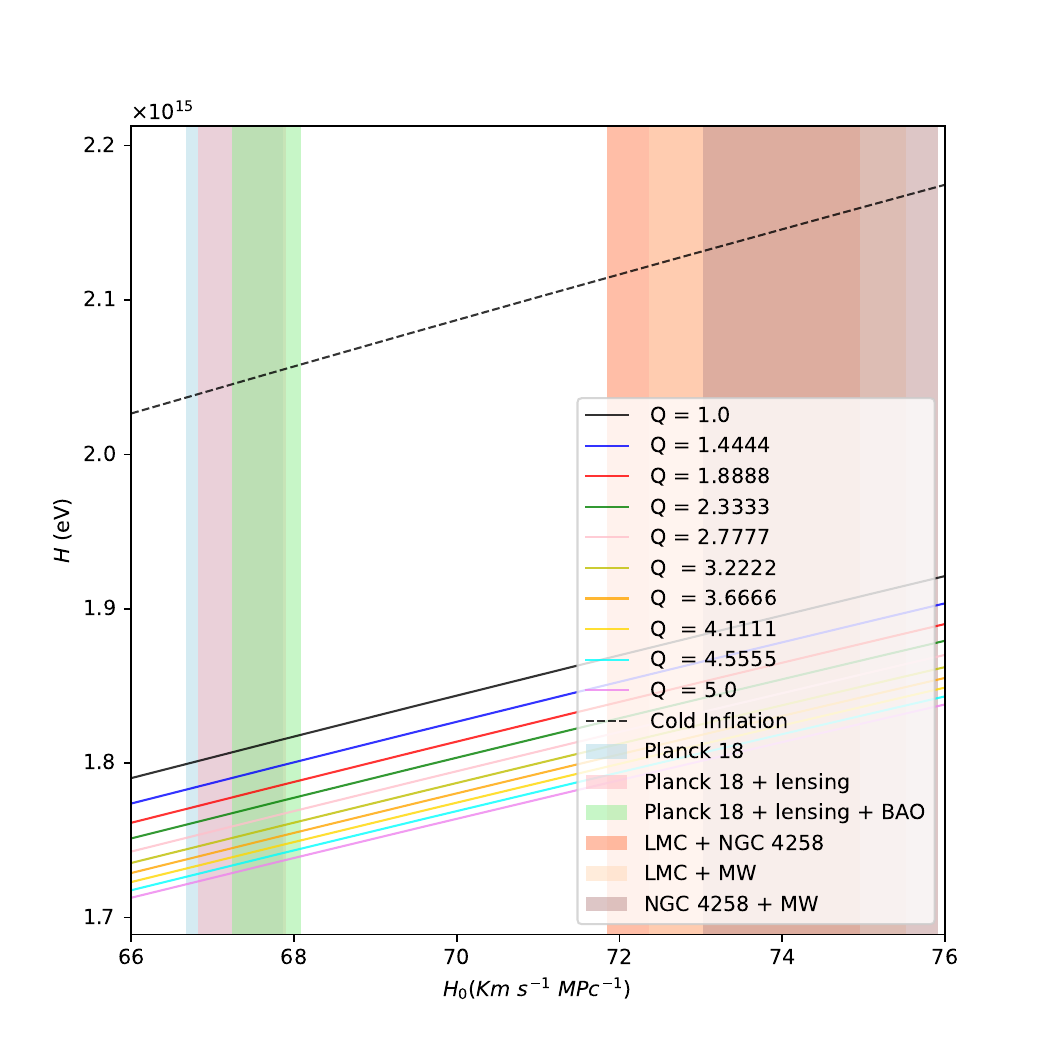}
     \caption{Quadratic}
     \end{subfigure}
     \begin{subfigure}{0.5\linewidth}
    \includegraphics[width=\linewidth]{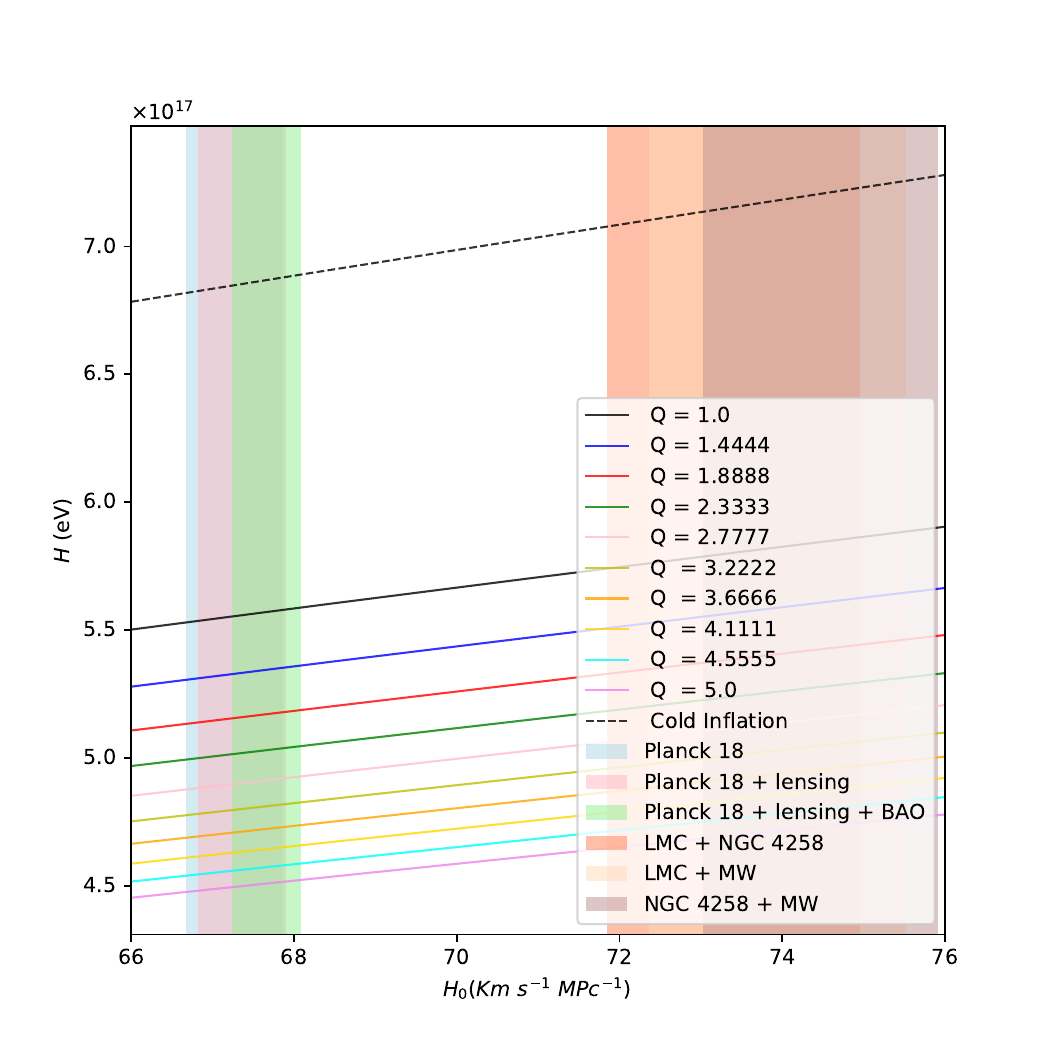}
    \caption{Quartic}
     \end{subfigure}
}\par
\makebox[\linewidth]
{
   \begin{subfigure}{0.5\linewidth}
   \includegraphics[width=\linewidth]{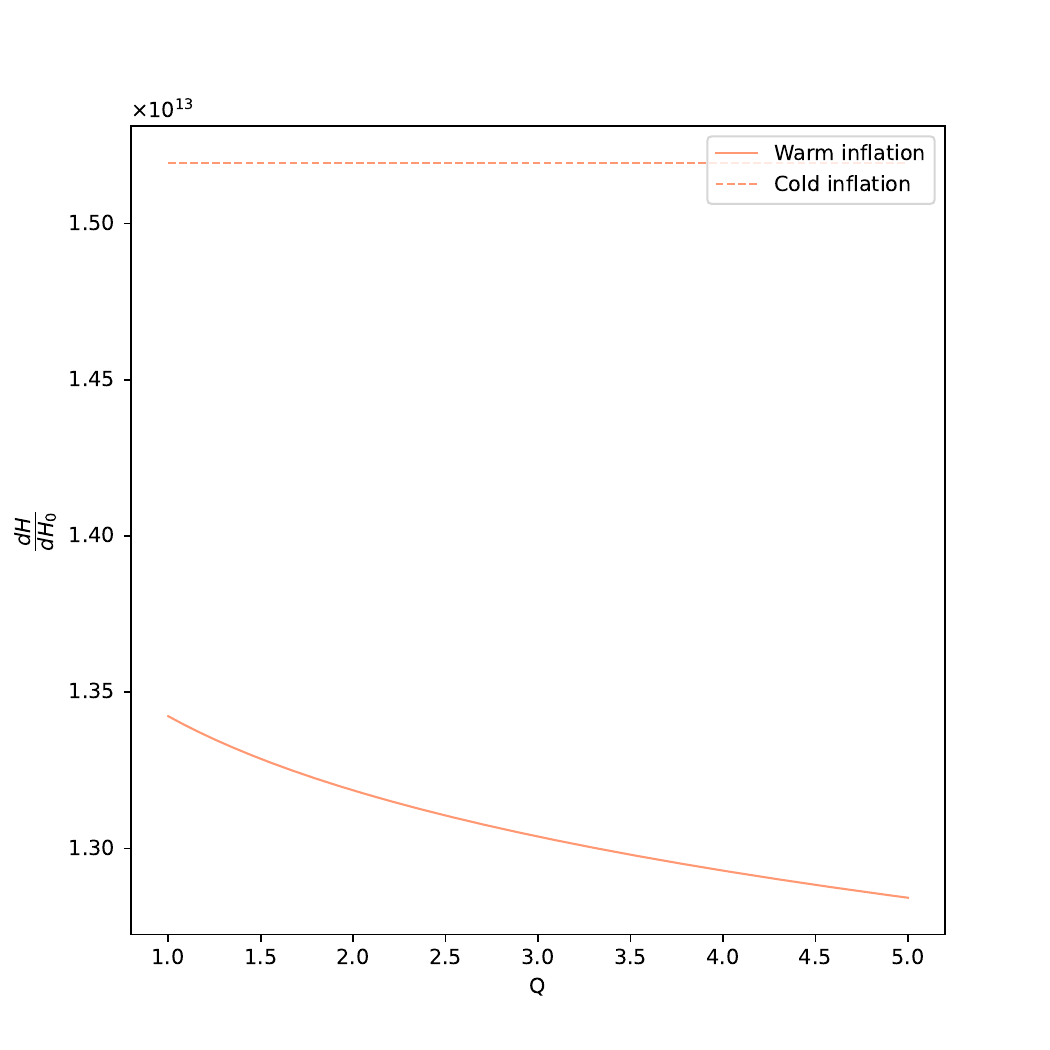}
     \caption{Quadratic}
     \end{subfigure}
    \begin{subfigure}{0.5\linewidth}
    \includegraphics[width=\linewidth]{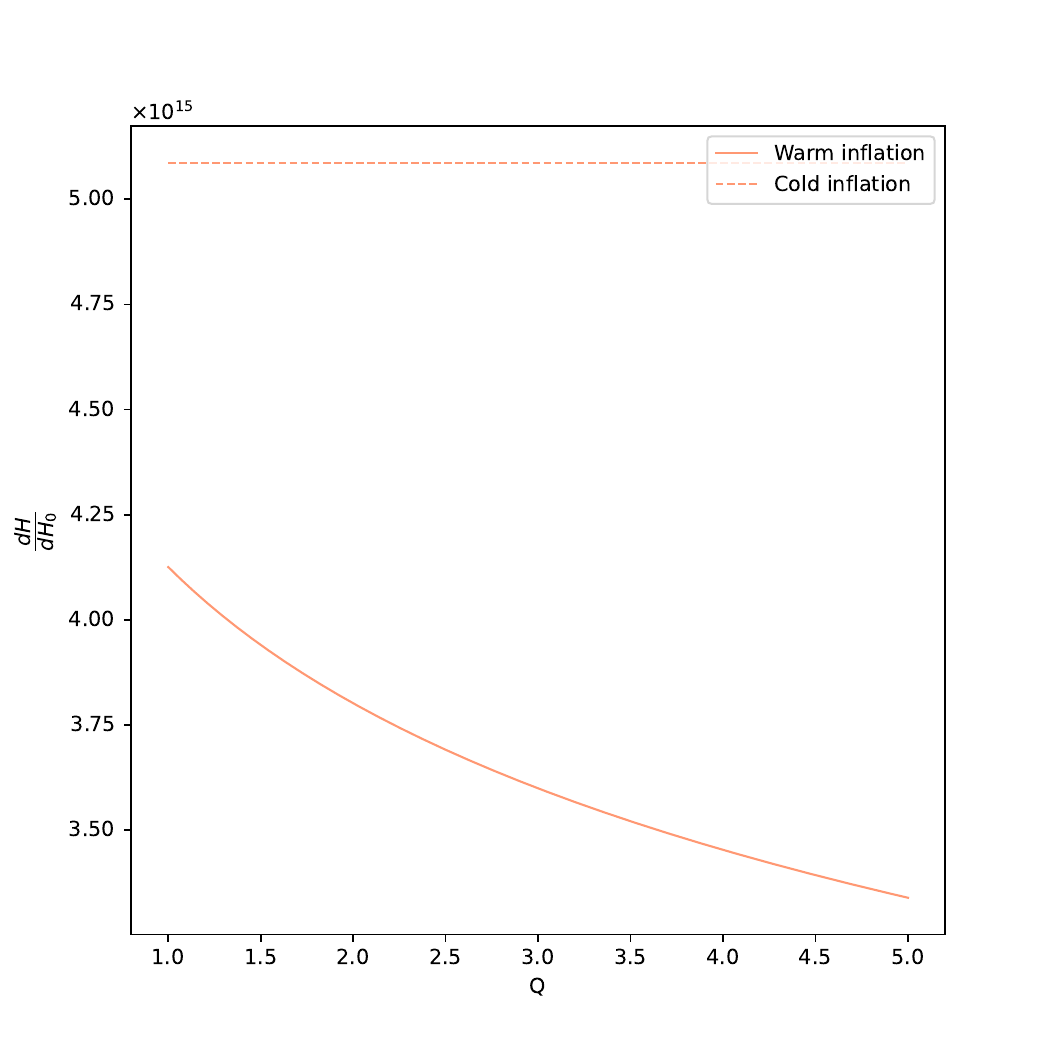}
    \caption{Quartic}
     \end{subfigure}
}
\caption{\label{HTDMM} Dissipation effect on the Hubble parameter during inflation (H) for a range of present Hubble parameter ($H_0$) for quadratic (a) and quartic (b) in DMM with Planck 2018 and SH0ES data (top panel).  The effect of Q on the slope ($s = \frac{dH}{dH_0}$ ) for quadratic (c) and quartic (d) in DMM with Planck 2018 and SH0ES data (bottom panel).}
\end{figure}
We examine the rate of variation of $H$ for different values of $Q$ using the Hubble parameter obtained from Planck 2018 and SH0ES data and the results are presented in Figures (\ref{HTDMM}, \ref{HTtwo}, \ref{HTwli}, \ref{HTax}). 
A close inspection of the results reveals a decrease in the slope associated with the Hubble parameter during warm inflation. The lowering of $s$ can be identified as a reduction in $dH$ due to the dissipative effect at the same time it can also be viewed as an increase in $dH_0$. Clearly, there is a distinction between the present Hubble parameter ($H_{0}$) in the cold inflation and the warm inflation, which becomes significant as the contribution from the dissipative sector increases. The lower value of $H_0$ can be linked to $\Lambda$CDM cold inflationary limit ($Q=0$) and  $H_0$ can take the higher value as suggested by the supernova based observations when warm inflation ($Q \neq 0$) is considered. Therefore both the higher and lower values of $H_0$ can be accounted. Candidate for WI can act like an evolving early dark energy that can explain the conflict between the lower values of $H_0$ from Planck 2018 and the higher values of $H_0$ measured by SH0ES. The latest observations from Dark Energy Spectroscopic Instrument (DESI DR2) \cite{desidr2} also confirms a time evolving dark energy equation of state \cite{giare} or an interacting dark energy \cite{supriya} that challenges the $\Lambda$CDM model. Recent works show that dark energy can be taken up for reconciling major tensions in cosmology \cite{odi, sami}. This suggests that Hubble tension may be reminiscent of thermal fluctuations seeded during warm inflation that is responsible for the dynamic nature of dark energy. The dominance of thermal fluctuations over quantum ($Q$) can be estimated from CMB, which makes this model observationally interesting. Based on the analysis of the obtained results, the rate of change of the Hubble parameter ($\frac{dH}{dH_0}$) in weak dissipative regimes (TSM and WLI) of quadratic inflation approaches and merges with that of the cold inflation at $Q=0$. Thus weak dissipative regime may be preferred over a strong dissipative regime for resolving Hubble tension. Recent work shows that DESI DR2 supports EFT formulation of dark energy \cite{qde} and quantum gravity \cite{qg}.  It is also known that quantum gravity can give a possible solution to Hubble tension \cite{anu}. It is interesting to note that the results of the present study also emphasise a quantum gravity phase before inflation. Therefore, we may conclude that the role of string theory motivated quantum gravity models of warm inflation is vital in interpreting the measured $H_0$ while unraveling the Hubble tension. 

\section{Discussion and conclusion}

The conventional cold inflationary scenario could not explain some of the issues like quantum to classical transition of primordial perturbation, requirement of a separate reheating phase and could not account for the observed lower tensor to scalar ratio. Warm inflation brought a radical change in understanding inflation by successfully addressing these issues through supersymmetry and string theory based models. Analysis of the four major models of warm inflation stemming from high energy theories like superstring theory with quadratic and quartic inflaton potential show that there is an enhancement in the scalar perturbations due to the interactions among the fields and its related dissipation, which leads to a decrease in the tensor to scalar ratio. Studies involving early universe models and their modifications struggle to give a satisfactory solution to the Hubble tension. Warm inflationary scenario can explain the disparity between the observed high and low values of the present Hubble parameter. Recently, it was shown that incorporating quantum gravity in hybrid inflationary model can provide a possible solution to the Hubble tension \cite{anu}. As the superstring theory is rooted from quantum gravity, the present study on the Hubble parameter with superstring theory based warm inflationary models complements these results.\\
\\
In the current study, we show that quantum gravity based superstring theory motivated models for thermal interactions in warm inflation can account for the two different values of the present Hubble parameter. The framework of warm inflation involves additional background fields that are analogous to multi field models such as in the case of hybrid inflationary model where inflaton drives the inflation and the waterfall field mediates the phase transition. The results of the present study also insist on the presence of multiple fields during the onset of inflation. This can be easily realised from the wide range of vacua available in the string landscape. Hubble tension can be addressed by both quantum gravity and warm inflation. Therefore, warm inflation may also be a reflection of quantum gravity. Multi fields are required to explain the variety of particles available in the present universe. Some of the fields can be a candidate for dark matter. There is further scope to explore the UV completion of the low energy models, which is beyond the scope of this work.\\
\\
The obtained scalar spectral index and tensor to scalar ratio for all the warm inflationary models are consistent with recent CMB observations. Among these different models of warm inflation, the minimal warm inflation where an axion couples to the non-abelian gauge field is found most suitable for achieving strong dissipation ($Q>>1$) with negligible backreaction and thermal corrections. Moreover, the upcoming CMB forecasts favours the strong dissipative regime of MWI. However, evolution of the Hubble parameter from lower to a higher value in warm inflation can be better visualised in the weak regime. The cold inflationary scenario can be recovered in the $Q \rightarrow 0$ limit of weak dissipative warm inflation and is favoured over the strong dissipative sector when compared with estimates from CMB.  Thus we conclude that warm inflationary scenario that accommodates quantum gravity and supersymmetry in a multi field framework can alleviate Hubble tension as well as provide a better understanding of the early universe, including inflation and particle production. Considering warm inflationary scenario is crucial in validating inflationary model and quantum gravity. The warm inflationary model successfully reinforces the requirement of multiple fields in the early universe and can be a classical test for quantum gravity.\\
\\

\section*{Acknowledgement}
AB acknowledges the financial support of Prime Minister’s Research Fellowship (PMRF ID : 3702550) provided by the Ministry of Education, Government of India.\\
\\

\section*{Appendix}
\renewcommand{\theequation}{A.\arabic{equation}}
\setcounter{equation}{0}
The following are the detailed explanations and steps leading to Eq.(60).\\
\\
The scalar field $\phi(x,t)$ in the early universe also consists of tiny fluctuations $\delta \phi(x,t)$ apart from the homogeneous part $\phi(t)$. Therefore, 
\begin{equation}
\phi(x,t) = \phi(t) + \delta \phi(x,t). 
\end{equation}
In cold inflation, the fluctuations of a generic massive scalar field are quantum in origin, whose magnitude and variance can be obtained \cite{bour} from 
\begin{equation}
\ddot \delta  \phi(x,t) + [k^2 + m^2]  \delta  \phi(x,t) = 0 
\end{equation}
as
\begin{equation}\label{bj90}
<|\delta \phi_k|^2 > \  \ \simeq \  \frac{3H_I^4}{8\pi^2m^2} \bigg[ 1 - e^{- \frac{2m^2N}{3H_I^2}} \bigg] ,
\end{equation}
where $H_I$ is the Hubble parameter during inflation, $k$ is the wavenumber for the mode that exits the horizon, $m$ is the mass of the scalar field $\phi(t)$ and $N$ is the e-folding number that amounts the duration of inflation.\\ 
Inflation occurs for a long period ($N$ is very high $\sim $ 60) therefore equation (\ref{bj90}) becomes
\begin{equation}\label{fluct}
<|\delta \phi_k|^2 >  \ \ \simeq \   \frac{3H_I^4}{8\pi^2m^2} .
\end{equation}
The inflationary potential can take the form,
\begin{equation}\label{orinf}
V(\phi) = \frac{1}{2}m^2 \phi^2 .
\end{equation}
The quantum fluctuations can dominate over homogenous part of the field in the early universe. Therefore using equation (\ref{fluct}) in (\ref{orinf}) the potential can be rewritten as
\begin{equation}\label{V}
V(\phi) \  \simeq \ \frac{1}{2}m^2  <|\delta \phi_k|^2 > \ \simeq   \ \frac{3H_I^4}{16\pi^2} .
\end{equation}
The Friedmann equation gives,
\begin{equation}\label{bj}
V(\phi) \  \simeq 3m_{pl}^2H^2.
\end{equation}
From Eq(\ref{V}) and (\ref{bj}),
\begin{equation}\label{st}
3m_{pl}^2H^2 = \frac{3H_I^4}{16\pi^2}.\\
\end{equation}
Also the Hubble parameter is defined as,
\begin{equation}
H^{2} = (\Omega_m+\Omega_r+\Omega_\Lambda ) H_{0}^{2}. \nonumber
\end{equation}
During the onset of inflation $\Omega_m+\Omega_r=0$.
Therefore, \begin{equation}\label{sc}
H^2_I = \Omega_\Lambda  H_0^2.
\end{equation}
Here, it is assumed that the early dark energy component evolves to become the present cosmological constant ($\Lambda$) with fractional energy density $\Omega_\Lambda$.
Substituting Eq(\ref{sc}) in Eq(\ref{st}) and rearranging, we get
\begin{equation}\label{bj21}
H_I = (\Omega_{\Lambda})^{\frac{1}{4} } \sqrt{4 \pi m_{pl} \ H_0 } .
\end{equation}
The relation between $H_I$ and $H_0$ is discussed in \cite{anu} and \cite{bour} exclusively for the case of cold inflation where the term $F(Q)$ is absent.\\
\\
In warm inflation, the fluctuations are thermal in origin and take the form (as given in Eq(\ref{pert})),
\begin{eqnarray}
<\delta \phi^2> \  \sim
\cases{
 \frac{3HT}{4\pi} & for WI with $Q<<1$ \cr \nonumber
\frac{\sqrt{\gamma H} T}{2\pi^2} &  for WI with $Q>>1$. \cr }\nonumber
\end{eqnarray}
Consider Minimal warm inflation ($Q>1$, $<\delta \phi^2> \  \sim \frac{\sqrt{\gamma H} T}{2\pi^2}$ )  with quadratic inflation.
Substitute, Eq(\ref{pert}) and Eq(\ref{tH}), in Eq(\ref{V}) and repeat the steps from Eq(\ref{V}) to (\ref{bj21}) to get, 
\begin{eqnarray}\label{quad}
H_I = \Bigg[ \frac{12 \pi \Omega_\Lambda}{1+\frac{\sqrt{3}cQ^\frac{1}{4}}{\pi}} \Bigg]^\frac{1}{4} \sqrt{m_{pl}H_0}.
\end{eqnarray}
Similarly, Minimal warm inflation can also be studied with the quartic potential $V(\phi) = \frac{\lambda}{4}\phi^4$, to obtain a relation between $H_I$ and $H_0$,
\begin{eqnarray}\label{quart}
H_I = \Bigg[ \frac{48 \Omega_\Lambda}{\lambda(1+\frac{3cQ^\frac{3}{4}}{\pi^2})} \Bigg]^\frac{1}{4} \sqrt{\pi m_{pl}H_0}.
\end{eqnarray}
Eq(\ref{quad}) and (\ref{quart}) can be generalised as 
\begin{eqnarray} 
H = \cases{
          \bigg( \frac{12 \pi \Omega_\Lambda}{F(Q)}\bigg)^\frac{1}{4} \sqrt{m_{pl}H_0}, & for  $V(\phi)=\frac{1}{2}m^2\phi^2$ \cr
          \bigg( \frac{48 \Omega_{\Lambda}}{\lambda F(Q) }\bigg)^\frac{1}{4} \sqrt{\pi m_{pl} H_0}, & for $ V(\phi)=\frac{ \lambda}{4}\phi^4$ }
\end{eqnarray}
where,
\begin{eqnarray} 
F(Q) = \cases{
         1+\frac{\sqrt{3}cQ^\frac{1}{4}}{\pi}, & for  $V(\phi)=\frac{1}{2}m^2\phi^2$ \cr
         1+\frac{3cQ^\frac{3}{4}}{\pi^2}, & for $ V(\phi)=\frac{ \lambda}{4}\phi^4.$ \cr }
\end{eqnarray}
The other models of warm inflation can be studied with quadratic and quartic inflation and the generic relation between the Hubble parameter during warm inflation ($H$) and the present Hubble parameter ($H_0$) is given in Eq(60) and the various form F(Q) may take in the different combinations are given in Table 6. \\
\\

\end{document}